\documentclass[acmsmall]{acmart}

\AtBeginDocument{%
  \providecommand\BibTeX{{%
    \normalfont B\kern-0.5em{\scshape i\kern-0.25em b}\kern-0.8em\TeX}}}

\setcopyright{rightsretained}
\acmJournal{PACMHCI}
\acmYear{2021} \acmVolume{5} \acmNumber{CSCW2} \acmArticle{428} \acmMonth{10} \acmPrice{}\acmDOI{10.1145/3479572}

\usepackage{multirow}
\usepackage{subcaption}
\usepackage{graphicx}
\usepackage{adjustbox}

\usepackage{xcolor}
\usepackage{pifont}
\usepackage{enumitem}
\usepackage{hhline}
\usepackage{booktabs}
\begin{document}

\title[The Impact of Algorithmic Risk Assessments on Human Predictions]{The Impact of 
%Recidivism 
Algorithmic Risk Assessments on Human Predictions and its Analysis via Crowdsourcing Studies}

\orcid{1234-5678-9012}
\author{Riccardo Fogliato}
% \authornotemark[1]
\email{rfogliat@andrew.cmu.edu}
\affiliation{%
  \institution{Carnegie Mellon University}
  \streetaddress{5000 Forbes Avenue}
  \city{Pittsburgh}
  \country{USA}
  \postcode{15213}
}

\author{Alexandra Chouldechova}
\affiliation{%
  \institution{Carnegie Mellon University}
  \streetaddress{5000 Forbes Avenue}
  \city{Pittsburgh}
  \country{USA}
  \postcode{15213}}
\email{achoulde@cmu.edu}

\author{Zachary Lipton}
\affiliation{%
  \institution{Carnegie Mellon University}
  \streetaddress{5000 Forbes Avenue}
  \city{Pittsburgh}
  \country{USA}
  \postcode{15213}}
\email{zlipton@cmu.edu}

%%
%% By default, the full list of authors will be used in the page
%% headers. Often, this list is too long, and will overlap
%% other information printed in the page headers. This command allows
%% the author to define a more concise list
%% of authors' names for this purpose.
\renewcommand{\shortauthors}{Riccardo Fogliato, Alexandra Chouldechova, \& Zachary Lipton}

%%
%% The abstract is a short summary of the work to be presented in the
%% article.
\begin{abstract}
As algorithmic risk assessment instruments (RAIs) are increasingly adopted to
assist decision makers, 
their predictive performance 
% and properties vis-a-vis fairness 
and potential to promote inequity
have come under scrutiny. 
However, while most studies examine these tools in isolation, 
researchers have come to recognize 
that assessing their impact 
requires understanding the behavior 
of their human interactants.
In this paper, building off of
several recent crowdsourcing works 
focused on criminal justice, 
we conduct a vignette study 
in which laypersons are tasked 
with predicting future re-arrests. 
Our key findings are as follows: 
(1) Participants often predict 
that an offender will be rearrested
even when they deem the likelihood of re-arrest 
to be well below 50\%; 
(2) Participants do not anchor on the RAI's predictions; 
(3) The time spent on the survey varies widely 
across participants and most cases 
are assessed in less than 10 seconds;
(4) Judicial decisions, unlike participants' predictions, 
depend in part on factors that are orthogonal 
to the likelihood of re-arrest. 
These results highlight the influence
of several crucial but often overlooked design decisions
and concerns around generalizability 
when constructing crowdsourcing studies
to analyze the impacts of RAIs.

\end{abstract} 

%%
%% The code below is generated by the tool at http://dl.acm.org/ccs.cfm.
%% Please copy and paste the code instead of the example below.
%%

% \ccsdesc[300]{Human-centered computing~Haptic devices}

%%
%% Keywords. The author(s) should pick words that accurately describe
%% the work being presented. Separate the keywords with commas.

%%
%% This command processes the author and affiliation and title
%% information and builds the first part of the formatted document.
\maketitle

\section{Introduction}
 % add references to Online panels in social science research: Expanding
 % sampling methods beyond Mechanical Turk by Chandler et al + others
Risk assessment instruments (RAIs) are increasingly deployed 
in critical domains, including finance, education, criminal justice, 
child welfare, hiring, and healthcare \cite{paravisini2013incentive, delen2010comparative, berk2018fairness, saxena2020human, sendak2020human, raghavan2020mitigating}.  
Rationales for adopting these tools often center around efficiency: 
The hope is that RAIs might offer fast, accurate, objective, 
and cheap decision-making at scale \cite{predalg20nyt}. 
To assess the potential benefits and disadvantages of RAIs, 
common practice is to compare their statistical properties against the status quo. 
However, while most research on algorithmic fairness, accountability,
and transparency has focused on RAIs in isolation, a broad recognition is
emerging, particularly within the HCI community, 
that many of these tools must be studied 
as situated in human-in-the-loop systems.

While it can be challenging to assess 
the predictive performance and fairness
properties of RAIs in isolation 
\cite{kleinberg2017human, kallus2019assessing, lakkaraju2017selective, fogliato2020fairness, coston2021characterizing}, investigating their impact on systems 
where the human is the ultimate decision maker 
introduces a new layer of complexity. 
Perhaps the most direct evidence bearing on these questions 
comes from longitudinal studies of real-world systems \cite{de2020case,chouldechova2018case}, 
particularly within the criminal justice setting. 
Several such studies have argued that these RAIs 
did not substantially improve the quality 
of the judicial decisions \cite{stevenson2018assessing,berk2017impact} 
and in some cases amplified the existing
disparities \cite{evidencealbright}. 
In studies that reported more positive outcomes, 
the RAI was often deployed as part of broader criminal justice reforms, 
making it difficult to isolate the effect of the algorithmic tool 
from other explanatory factors \citep{grant2018criminal, austin2020bail}.

In order to overcome the difficulties 
of characterizing decision-making 
and the impact of RAIs on systems 
where the human is in-the-loop,
many researchers have turned to lab experiments 
\cite{yin2019understanding, poursabzi2021manipulating,zhang2020effect, bansal2021does}. 
Findings from these studies, however, may not generalize 
to real-world high-stakes contexts because study participants, 
who are typically recruited on crowdsourcing platforms,
are not representative of experts and because, unlike experts, 
they are not actually making decisions (only predictions). 
Still, results from these studies can elucidate general
phenomena that characterize human-in-the-loop decision-making frameworks. 
In addition, the predictive performance achieved by study participants 
can arguably be seen as a benchmark for expert decision makers. 

Many studies in this line of research have focused on the criminal justice context 
\citep{dressel2018accuracy, biswas2020role, green2019disparate, grgic2019human, green2019principles, jung2020limits, mallari2020look}. 
These works have mainly examined predictive performance and fairness properties 
concerning the study participants' predictions of defendants' future criminal recidivism, 
often using the ProPublica COMPAS dataset \cite{angwin2016machine}.
However, their survey designs exhibit several key differences,
such as as the number and type of predictions elicited from participants, 
the structure of financial rewards, 
and the requirements participants had to satisfy 
to take part in the study (longer discussion in \textsection\ref{sec:related_work}). 

In this paper, we first characterize 
some of the aforementioned differences
across survey designs 
and then introduce our empirical study 
to determine how these factors impact 
crowdworkers' predictions of future criminal recidivism.
More specifically, we identify four research questions (RQs) 
that we believe are important considerations 
for assessing the generalizability of results
from these studies but, to our knowledge, 
have not been addressed by prior work:
\begin{description}[topsep=0pt]
    \item[RQ1] Do evaluations of participants' responses depend on the type
    of predictions that are elicited? More specifically, can we infer participants'
    (binary) outcome predictions only based on their (probability) risk
    estimates? Does this matter for evaluations of participants' predictive
    performance and reliance on the RAI?%only an estimate of the risk that is
    %then converted into a binary prediction by the researcher
    %Then, does this matter for evaluations of participants' reliance on the
    %RAI? %How do participants integrate the RAI's recommendations into their
    %decision-making process? In other words, how do participants revise their
    %risk estimates and binary predictions after seeing the RAI's?
    \item[RQ2] What is the impact of anchoring effects on participants'
    predictions? In other words, do participants respond differently if they are
    asked to pre-register provisional responses before being shown the RAI's
    recommendation?
    \item[RQ3] How much time do participants spend on the assessments?
    %i.e., how much time do they spend on each assessment? % or say something
    %with satisficing
    \item[RQ4] How do predictions (of re-arrest) and judicial decisions (to
    incarcerate) differ?
\end{description}

\noindent To investigate RQ1--4, 
we designed a vignette study (the ``survey'')
where laypersons recruited on Amazon Mechanical Turk 
were asked to predict future recidivism 
of a series of criminal offenders with and without the
assistance of an RAI (see~\textsection\ref{sec:data_methods}). 
Participants were shown short descriptions of the offenders
and then asked to assess the likelihood and predict 
whether the given offenders would be rearrested. 
By employing a between-subjects design, 
we tested whether participants anchored on the RAI's recommendations 
when these were presented at the outset 
together with the offender's description.

Our results (\textsection\ref{sec:results}) indicate 
the importance of asking participants separately
about their \textit{probability} predictions 
and their (binary) \textit{outcome} predictions: 
{\bf (1)} The study participants often predicted re-arrest 
even when they deemed the likelihood of re-arrest to be well below 50\%.
%(see~\textsection\ref{sec:mapping_prob_bin}).
The distinction matters for evaluations 
of participants' predictive performance
and reliance on the RAI. 
For example, we find that participants 
substantially updated their risk estimates 
in the direction of the RAI's recommendation,
but they rarely revised their binary predictions 
to match the RAI's (binary) prediction. 
Interestingly, participants self-reported 
revising their risk estimates almost twice 
as often as their binary predictions 
after seeing the RAI's recommendation.
{\bf (2)} Interestingly, we do not find any evidence 
of participants anchoring on the RAI's predictions. 
Instead, surprisingly, the revised risk estimates 
made by participants in our non-anchoring condition 
were significantly {\it closer} to the RAI's.
{\bf (3)} By tracking the time spent on each vignette,
we discover that many participants take a long time to complete the survey 
but actually spend surprisingly little time on each case 
(average=15 seconds, median=10), taking long breaks.
%(\textsection\ref{sec:time}).
Moreover, because many participants perform similarly and few outperform the RAI, 
their predictive accuracy does not constitute a reliable proxy of time spent. 
{\bf (4)} Finally, we discuss crucial differences
between the predictive task and judicial decisions
that make it difficult to draw direct conclusions
about the latter from crowdsourcing studies such as ours. 
% show that the impact of RAIs on participants' predictions does not have direct
% and obvious implications in terms of judicial decision making. This is mainly
% due to the fact that decisions are correlated with 
We present an analysis of real judicial decisions, 
which suggests that judicial decisions, 
unlike participants' predictions,
depend not only on the offender's likelihood of recidivism 
but also on the gravity of the crimes committed.

To facilitate the reproducibility of our analysis, 
we have obtained IRB approval from Carnegie Mellon University 
and participants' consent to publicly share 
the data collected as part of our experiment. 
Our data and code for the analysis are available at 
\url{www.github.com/ricfog/the-impact-of-algorithmic-rais}.

\section{Background}\label{sec:related_work}
%\noindent \textbf{Predicting criminal recidivism: 
%comparing humans, RAIs, and RAI-assisted humans via lab experiments}

%Rather than addressing real decisions, 
%lab experiments typically focus on crowdsourcing studies,
%where participants are recruited 
%on online crowdsourcing marketplaces---mainly 
%Mechanical Turk---and have little or no domain knowledge.
% In this section, 
We focus our treatment of related work
primarily on crowdsourcing studies 
on the prediction of criminal recidivism~\citep{dressel2018accuracy, biswas2020role, tan2018investigating, green2019disparate, grgic2019human, green2019principles, jung2020limits, mallari2020look}. 
These studies mainly involve 
related prediction tasks
in which participants are asked 
to predict the likelihood that a defendant or an offender, 
if released, will be rearrested 
within a specified period of time. 
Factors that would likely influence 
decisions but not predictions,
such as the defendant's ability to pay bail 
or their culpability, 
are intentionally excluded.
Factors that are not recorded 
in the data used by the RAI,
such as the defendant's and prosecutor's arguments, 
are also (necessarily) left out.
These studies speak to the gains in efficiency 
that should be expected 
if judges were to make decisions only based 
on the defendant's likelihood of re-arrest 
and the information present in the data. 
As our analysis will show (see~\textsection\ref{sec:pred_vs_dec}), 
these results do not directly inform 
the impact of future deployments of RAIs; 
yet, they potentially provide a benchmark 
for the \textit{predictive} performance 
of these human-in-the-loop systems.
%systems where the human is in-the-loop.

Existing studies have mainly focused on two key questions:
(i) how the predictions of study participants (alone) compare to the RAI's;
and (ii) whether and how the RAI's recommendations are taken into account by participants. 
In terms of the former,
the influential study of \citet{dressel2018accuracy} 
found that, on the ProPublica COMPAS dataset~\citep{angwin2016machine},
the predictive accuracy of their study participants' (crowdworkers)
was comparable to that of the COMPAS RAI.
The authors concluded that laypersons
performed no worse than COMPAS.
Successive studies replicated this finding~\citep{jung2020limits, mallari2020look, grgic2019human}. 
However, they also found that participants' predictive performance 
was considerably lower than the RAI's 
when outcome feedback was removed, 
the base rate was decreased,
or when the area under the curve (AUC), 
instead of accuracy, was considered~\cite{jung2020limits}. 
These results should not be surprising for two reasons. 
First, humans tend to ignore information about the base rate, 
a phenomenon known as ``base rate neglect''~\cite{o2006uncertain, koehler1996base}. 
This bias, however, is mitigated when outcome feedback is provided \cite{gluck1988conditioning}.
Second, even simple RAIs achieve predictive accuracy 
similar to that of COMPAS \cite{angelino2017learning}, 
but ranking offenders by their risk of recidivism (i.e., regression) represents a ``harder'' statistical problem.
Studies in the second line of work, 
which investigated human predictions 
in presence of the RAI, 
generally found that the RAI led to small or no increments
in the predictive performance of participants' predictions \cite{vaccaro2019effects, green2019principles}.
Two related studies observed
differential compliance with 
the RAI's recommendations 
across defendants' racial groups:
They found that providing participants with the RAI's prediction 
led to a larger increase in the predicted risk of re-arrest for 
Black defendants 
compared to White defendants~\citep{green2019disparate, green2019principles}, a phenomenon that the authors named 
``disparate interactions''.
In our study, we found that participants, 
when presented with the RAI's recommendations, 
performed worse than the RAI 
across all the metrics that we considered (\textsection\ref{sec:performance}). 
In addition, we note that we were not able 
to replicate the disparate-interactions effect
(\textsection\ref{sec:race}).

However, some of the survey designs employed 
by these experiments present notable differences. 
In the following paragraphs, 
we describe some of such differences and discuss, 
based on both past and our own work, 
how these design choices 
% could have impacted the
can impact the predictions made by study participants.

In eliciting predictions, some studies asked participants for their risk
estimates (probabilities) \citep{green2019disparate, jung2020limits}, whereas
others asked for (binary) outcome predictions of
re-arrest~\citep{dressel2018accuracy, mallari2020look}.
\citet{vaccaro2019effects} collected both and \citet{grgic2019human} had
participants provide confidence ratings. To derive binary predictions from risk
estimates, \citet{jung2020limits} assumed that participants \emph{would have}
predicted re-arrest for a defendant whenever their risk estimate was larger than
50\%. However, it is well known that many individuals, even expert decision
makers, can struggle with poor numeracy skills~\cite{black1995perceptions,
schwartz1997role, lipkus2001general, visschers2009probability}. In addition,
individuals may base their decisions on a distorted version of the estimate of
the risk, e.g., by overweighting small probabilities~\cite{kahneman2013prospect,
prelec1998probability}. One recent crowdsourcing study on uncertainty
visualization has found that individuals that were poor at probability judgments
did well on decision tasks~\cite{kale2020visual}. Consistent with this evidence,
our results support the hypothesis that participants' binary predictions cannot
be simply obtained by converting risk estimates at a fixed threshold
(\textsection\ref{sec:mapping_prob_bin}). 
% For example, we find that our study participants 
% revised their risk estimates 
% in the direction of the RAI's, 
% but rarely changed their binary predictions (\textsection\ref{sec:nonanc}), and not always in the direction of the RAI's.

Past studies also differ in the stage 
at which the RAI's recommendation was presented
to the participants. 
In one of these studies \cite{green2019disparate}, 
the defendant's description and the RAI's recommendation 
were presented to participants at the same time. 
In another \cite{grgic2019human}, 
the RAI's information was made available 
only after the participant had made an initial prediction. 
However, the psychology and behavioral economics literature suggest
that participants might rely on the RAI more heavily 
in the former setting due to a phenomenon 
known as the {\it anchoring effect}, a.k.a. ``anchoring and adjustment heuristic''~\cite{o2006uncertain, tversky1974judgment, epley2005effortful}.
% 
% ZACK: Can probably tighten this up
% 
% also i don't know if "paradigm" is the right word here
According to this heuristic, participants that are presented with a novel
``anchor'' would first try to assess whether the anchor equals the target and
then adjust its value. The final answer generally tends to be biased in the
direction of the anchor. In the setting of our study, anchor and target are
represented by the RAI's recommendation and the defendant's recidivism outcome
respectively. Since individuals are often unable to accurately describe their own
cognitive mental processes~\cite{nisbett1977telling, wilson2004strangers},
hypotheses around this phenomenon need to be empirically tested. 
%In a study similar to ours, \citet{green2019principles} found that participants that were directly shown the RAI achieved lower accuracy than their counterparts.
This cognitive bias has been shown to affect judicial decision-making~\cite{enough2001sentencing,englich2005last,englich2006playing, mussweiler2000numeric, bushway2012sentencing}
and it has been studied
%in a real-world deployment in the context of lending~\cite{paravisini2013incentive}
in at least three other crowdsourcing experiments \cite{vaccaro2019effects, green2019principles, buccinca2021trust}, two of which focused on recidivism RAIs.
% Close to our study is
Close to our work is \citet{green2019principles}, which examined the effect of
anchoring on participants' risk estimates, finding that participants that had pre-registered their answers achieved
higher predictive performance. In our experiment, performance was nearly
identical across the two settings. Surprisingly, we found no evidence of
anchoring bias: The risk estimates made by the participants that had to
pre-register their predictions before seeing the RAI's recommendations were
closer to the predictions made by the RAI (\textsection\ref{sec:study_anc}).

The studies also differ considerably %present differences
in the number of defendants' profiles that were shown,
the requirements 
% (when needed) 
that workers had to satisfy in order to participate, 
the structure of the financial compensation, 
the number and form of attention checks.
These factors likely impacted the amount of effort 
made by participants in the study.
But none of the studies has tried to quantify 
the effort that participants make on each prediction, 
an aspect that seems 
% overwhelmingly
especially 
important for
ecological validity.
Problematically, 
% in the kind of tasks employed by these experiments, 
for these tasks,
even participants that answer carefully
can achieve low predictive accuracy 
and thus 
% such a metric is unlikely to represent 
accuracy
may not be a
% valid
reliable proxy for effort. 
% An alternative metric that could be 
Alternatively, one might consider 
the time spent by participants on the entire survey, 
which many past studies have reported~\cite{green2019disparate, vaccaro2019effects, mallari2020look, grgic2019human}. 
By taking into account the number of defendants 
shown in each of these surveys, 
a quick computation suggests that, 
on average, the prediction for a single defendant 
may take less than 15 seconds.
In our study, we show that, 
while this actually corresponds 
to the average time spent by our survey participants on each of the assessments, it varies widely both across participants 
and across vignettes (median=10 seconds)
(\textsection\ref{sec:time}). 

Another difference in survey designs that is worth mentioning, but whose
consequences we don't explore in this work, is the type of the RAI's
recommendation that was displayed in the vignette. In past studies, participants were presented
with the raw risk estimate of the RAI~\cite{green2019disparate}, the prediction
binned into risk scores 
%(e.g., 4:low score vs 7:high score)~
\citep{vaccaro2019effects}, or just the binary
prediction~\cite{grgic2019human}. The type of RAI's prediction that is
communicated has been shown to impact participants' judgments in crowdsourcing
studies~\cite{lai2019human, zhang2020effect} and even decisions made in the
criminal justice context~\cite{scurich2011effect}. In
\textsection\ref{sec:mapping_prob_bin}, we show that participants often do not
revise their binary prediction in the direction of the RAI's. In contrast,
\citet{grgic2019human} found that when study participants revised their (binary)
predictions, they did so to match the RAI's in almost the majority of cases. In
light of our results around how people convert risk estimates into binary
predictions, the seeming contrast between the two results is likely attributable
to our design choice of providing participants only with the RAI's risk estimate
and not its binary prediction.

%\subsection{Predictions of criminal recidivism and judicial decisions of whether to incarcerate}
We now focus on one last important difference: 
the target of the studies.
As we mentioned before, 
the goal of some of these experiments 
was to compare the predictive performance of the RAI 
with that of laypersons~\cite{dressel2018accuracy,jung2020limits}.  
The goal of other studies was, instead, 
to highlight potential unintended consequences 
% of the impact of
arising from the use of RAIs 
in judicial decision-making~\cite{grgic2019human, green2019disparate}.
However, there is a clear disconnect between 
{\it predictions} of re-arrest, 
which were collected through the surveys, 
and {\it decisions} (e.g., of whether to incarcerate), 
which are made by judges.
In particular, the risk of criminal recidivism, 
on which participants' predictions are based, 
represents only one of multiple factors 
that judges (are allowed to) account for
in their decision-making process. 
Our results show that these differences 
cannot be easily reconciled
% at least for the setting of criminal sentencing
in the criminal sentencing setting, 
especially in context of sentencing considered in our study
(\textsection\ref{sec:pred_vs_dec}).

% also the type of scale used matters Violence Risk Assessment and Risk Communication: The Effects of Using Actual Cases, Providing Instruction, and Employing Probability Versus Frequency Formats

\section{Data and methods}\label{sec:data_methods}

This section is organized as follows. First, we provide an overview of the
dataset and of the statistical modeling used to develop the RAI
(\textsection\ref{sec:data_rai}). We then cover task and experimental
design (\textsection\ref{sec:survey_design}), procedure
(\textsection\ref{sec:procedure}), and recruitment process
(\textsection\ref{sec:recruitment}). Lastly, we present the methodology used for
the analysis of the results (\textsection\ref{sec:data_analysis}). In Appendix
\textsection\ref{sec:comparison_pilot}, we also describe a small pilot study
that we ran while developing the current experiment. The pilot is different
in several ways from the main study, such as in the compensation scheme, which
was not tied to performance.

\begin{table}[t]
\centering
\begin{tabular}{lll}
\textbf{} & \textbf{dataset} & \textbf{survey sample} \\ 
% \hhline{===}
\toprule
share of male offenders & 79.0\% &  79.1\% \\
% \midrule
share of White offenders & 65.3\% & 65.1\% \\
% \midrule
number of prior arrests (sd) & 3.8 (4.9) & 3.8 (4.9)\\ 
% \midrule
mean age (sd) & 31.2 (10.4) & 31.4 (10.3)\\ 
% \midrule
mean age of first arrest (sd) & 23.5 (8.6) & 23.5 (8.6) \\
%\midrule
prevalence of re-arrests & 41.9\% & 42.0\% \\ 
%\midrule
number of offenders & 117464 & 3523 \\
\bottomrule
\end{tabular}
\caption{Summary statistics for the offenders 
contained in the full dataset and in 
the subset of offenders extracted for the survey. 
Age is measured in years. Standard deviations (``sd'') are reported in parantheses.
The marginal distributions of the 
offenders' prior number of charges, type of offense committed, 
and age of first arrest in dataset and survey sample are also similar, but are omitted from the table. 
}\label{tab:sample}
\end{table}

\subsection{Data and development of RAI}\label{sec:data_rai}

\subsubsection{Data}\label{sec:data}

The set of offenders used in our survey comes from a private dataset provided by 
the Pennsylvania Commission on Sentencing.  
The data contain information about offenders sentenced 
in the state's criminal courts. 
In our analysis, we considered only offenders whose race, 
as recorded in the data, was White or Black. 
The final dataset consisted of 117,464 observations, 65\% of which corresponded to White offenders. 
For each of the offenders, we know whether they were rearrested
within three years from the time of release from prison 
or imposition of community supervision. 
The overall base rate was 41.9\%, while the re-arrest rates for Black and White offenders were 51\% and 37.1\%, respectively.
We split the full data into train and test sets (70\%-30\%) prior to model training. 
A sample of 3,523 observations were further selected from the test set
using stratified random sampling to ensure 
that the resulting sample reflected the test population 
on race, sex, age, and re-arrest status.
These observations were used in the survey.  
Summary statistics for the dataset and the survey sample are reported in Table~\ref{tab:sample}.

\subsubsection{Development of the RAI}\label{sec:stat_model} % 

We used the available data to train RAIs that predict 3-year post-release
re-arrest using demographic features (age, sex, and race),\footnote{We included
race among the predictors because all modeling approaches on our data that
excluded this feature resulted in re-arrest risk being overestimated for White
offenders relatively to Black offenders. This over-estimation phenomenon has
been documented for other RAIs, including some presently in use or under
consideration for use~\cite{pcsriskass, demichele2018public}. The use of race as
an input to improve the predictive bias properties of the model has recently
been discussed in~\citet{skeem2020using}. Here, we acknowledge one strong
objection to the use of risk assessment in criminal justice decision-making:
Re-arrest may be a racially biased measure of offending---due to disparities in
how different groups are policed. Thus an unbiased predictor of re-arrest may
nevertheless be a biased predictor of offending \citep{fogliato2020fairness}.
However, the participants in our study are explicitly asked to assess the
likelihood of re-arrest, so the RAIs predictive bias as a predictor of re-arrest
is precisely the property at question.} information about the current charge
(type of offense and whether it is a misdemeanor or a felony), and several
features reflecting prior criminal history (e.g., past number of arrest and
charges for several offense categories). We trained logistic regression,
Lasso~\cite{tibshirani1996regression}, random forest~\cite{breiman2001random},
and extreme gradient boosting (XGBoost)~\cite{chen2016xgboost} models on the
training data, tuning the hyperparameters via cross-validation. The four models
showed similar performance on the holdout set: Prediction accuracy was around
66\% and the area under the curve (AUC) was approximately 70\%. These AUCs are
comparable to those of other recidivism RAIs that are deployed in jurisdictions
across the country, such as COMPAS \cite{dieterich2016compas} and the Public
Safety Assessment (PSA) \cite{demichele2018public}, among others
\cite{desmarais2020predictive}. All models were fairly well calibrated overall
and, at a classification threshold of 0.5, they all predicted re-arrest for
approximately 30\% of the offenders in the holdout set. We also assessed the
models for racial predictive bias, finding that the Lasso and random forest
models were both racially well-calibrated. More details on our assessment of
predictive bias in the predictions are provided
in~\textsection\ref{sec:app_racial_bias}. Given the similarity in overall
performance across the four models, we decided to adopt the Lasso for our
experiments, since it was the simplest model that we found to be racially
well-calibrated. For this part of the analysis, we used the
\texttt{tidyverse}~\cite{tidyverse} and~\texttt{tidymodels}~\cite{tidymodels}
sets of packages in \texttt{R}~\cite{Rcore}.

\subsection{Task and experimental design}\label{sec:survey_design} % 

\begin{figure}
%\centering
%\begin{subfigure}{.5\textwidth}
  \centering
  \includegraphics[scale = 0.3]{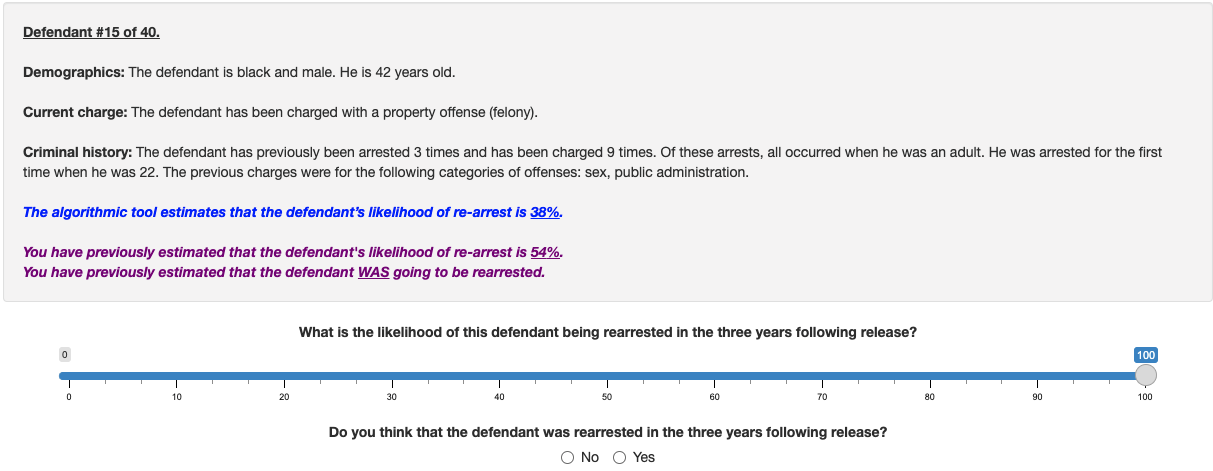}
  \includegraphics[scale = 0.3]{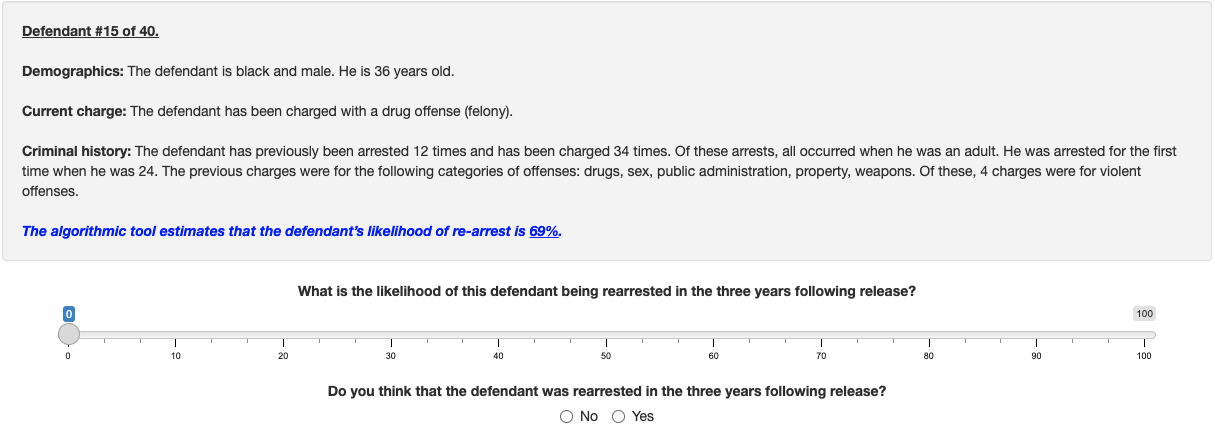}
%\end{subfigure}%
%\begin{subfigure}{.5\textwidth}
%  \centering
%  \includegraphics[width=\linewidth]{Figures/screen_anc.%png}
 % \caption{}
%\end{subfigure}%
\caption{The top and bottom panels provide examples of vignettes that were presented to participants in the non-anchoring and anchoring settings respectively. 
Participants assigned to the former setting were first asked to pre-register provisional responses that they were allowed to revise after being shown the RAI's recommendation (example here). 
Participants assigned to the anchoring setting were shown directly the RAI's recommendation. 
Both the likelihood estimate and the binary prediction were required to proceed to the following vignette. 
The initial value of the slider was randomly set at either 0\% or 100\%. 
}\label{fig:vignette_screenshot}
\end{figure}

\begin{figure}[t]
  \centering
  \includegraphics[width=\linewidth]{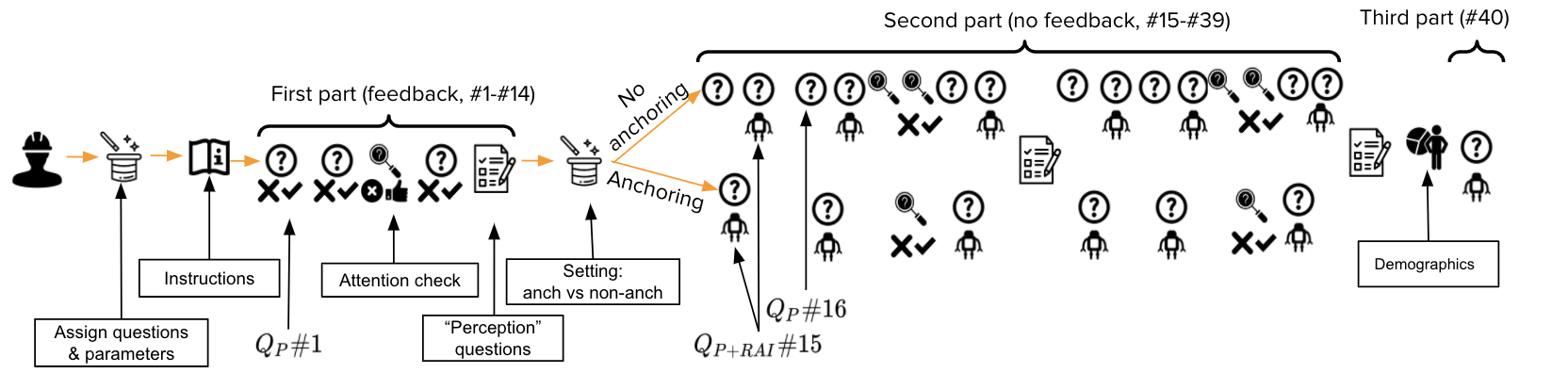}
\caption{Structure of the survey. $Q_{P+RAI}\#i$ and $Q_P\#i$
indicate the answers given by the participant with and without 
the assistance of the RAI, respectively, for the $i^{th}$ offender. 
At the initialization of the web app, 
the offender's profiles and the other parameters were chosen.
Then the participant went through the instructions (see also~\textsection\ref{sec:app_surveystructure}). 
In the first part of the survey, 
consisting of 14 offenders plus one attention check, 
outcome feedback was provided. 
At the end, the participant completed the ``perception questions'' 
and then proceeded to the second part of the survey.
Here, the participant did not receive feedback and was assigned to either 
the anchoring or to the non-anchoring setting. 
This part consisted of 25 offenders (\#15-39) plus two attention checks
and contained two sets of perception questions. 
Demographics were collected right before 
the third and last part of the survey, 
which consisted of only one question (\# 40). 
%Note that the condition was assigned 
%before the initialization of the app 
%(i.e., all participants in the same batch of surveys 
%published on Amazon Mechanical Turk were assigned to the same condition),
%whereas the setting was assigned only after 
%the participant had logged into the survey
%(i.e., participants within in the same batch 
%could be assigned to different conditions).
}
\label{fig:surveystructure}
\end{figure}

In the survey, participants were tasked with predicting the likelihood and the
occurrence of a re-arrest for a series of 40 offenders. The profile of each
offender was presented through a descriptive paragraph based on a subset of the
features that had been used to train the RAI.\footnote{The accuracy of the
models trained on this subset of features were all around 65\%-66\%, which is
nearly identical to the 66\% accuracy achieved by the Lasso model selected as
our final RAI. The additional features relied upon by the Lasso model were
helpful in achieving model calibration, but were excluded from the vignettes to
make the offenders' descriptions sufficiently brief for participants to process.
Those features are counts of the offender's past arrests for specific crime
types.} For each offender, participants were asked the two following questions:
($Q^p$) "What is the likelihood of this offender being rearrested in the three
years following release?" and ($Q^b$) "Do you think that the offender was
rearrested in the three years following release?".\footnote{As shown in Figure
\ref{fig:vignette_screenshot}, in the survey we used the term ``defendant''
rather than ``offender''.} The possible answers to $Q^p$ were probabilities on a
scale 0\%-100\% in bins of $1\%$ which could be selected using a slider scale.
The initial value of such slider was randomly set at either 0\% or 100\% when
the participant entered the survey. Instead, $Q^b$ required a negative or
positive answer which could be selected using radio buttons. Two examples of
vignettes are shown in Figure~\ref{fig:vignette_screenshot} (see also the
structure of the vignette in~\textsection\ref{sec:structure_vignette}). The core
task in the survey consisted of three consecutive parts, which consisted of
$14$, $25$, and $1$ offenders' assessments respectively (see
Figure~\ref{fig:surveystructure}). We now describe each of these in turn.

%An example of vignette is shown in Figure~\ref{fig:vignettes}. 

%\subsubsection{Structure of the survey}

In the first part of the survey (offenders \#1-14 in
Figure~\ref{fig:surveystructure}), participants were provided with outcome
feedback, but the RAI's predictions were not made available to them.
Participants were asked to record their responses ($Q^p_P$ and $Q_P^b$) and given feedback after each prediction. The feedback was as follows:
``The offender was/was not rearrested in the three years following
release'', which was shown in green if the participant's prediction for $Q^b$,
the binary prediction question, was accurate and in red otherwise.\footnote{In
these 14 trials, we provided outcome feedback to make participants more
comfortable with the task and, at the same time, to mitigate the possibility of
base rate neglect. However, it is likely participants learnt even after the end
of this part of the survey. For example, \citet{jung2020limits} reported using
only the last 10 (out of 50) responses given by participants because of learning
effects.} Offenders' profiles were randomly drawn from the survey sample.

In the second part of the survey (offenders \#15-39), outcome feedback was
removed but participants were shown the RAI's predictions. We employed a
between-subjects design to study the effect of anchoring, assigning participants
to one of two conditions, which we call \emph{anchoring} and \emph{non-anchoring}.
The participants that were assigned to the anchoring condition were shown the
offender's profile and the prediction of the RAI in the same vignette ($Q^p_{P+RAI}$ and $Q_{P+RAI}^b$). Here, the
RAI serves as the anchor. Instead, if assigned to the non-anchoring
setting, participants were first asked to estimate the likelihood ($Q_P^p$) and
predict the occurrence ($Q_P^b$) of a re-arrest based on the offender's profile
alone, as they had done during the first part of the study. After submitting a
response, they were presented with the RAI's prediction for the given case and
were allowed to revise their own predictions ($Q^p_{P+RAI}$ and $Q_{P+RAI}^b$).
The randomization scheme was based on Efron's biased coin
design~\cite{efron1971forcing}.\footnote{More specifically, the probability of
assignment was 1/2 for the first participant of the survey, and was adjusted
according to Efron's biased coin design~\cite{efron1971forcing} for successive
participants. According to this biased coin design, participants were assigned
to the anchoring setting with probability 2/3 if the majority of past
participants had been assigned to the non-anchoring setting, 1/3 if the majority
of past participants had been assigned to the anchoring setting, and 1/2
otherwise.} In this part of the survey, participants were shown the descriptions
of a set of offenders that were drawn from the survey sample either randomly or
controlling for the RAI's predictive accuracy to be around 67\%. However, due to
a glitch, some of the participants were shown all cases for which the RAI's
binary predictions were accurate first. One past work found that the order in
which the offenders' profiles were ordered did not affect participants' reliance
on the RAI, even when outcome feedback was given~\cite{grgic2019human}. Since
the study participants affected by the glitch were equally split across the
anchoring and non-anchoring conditions (see Table~\ref{tab:random}
in~\textsection\ref{sec:order_effect})  
and we could not detect any interaction of the two effects in some preliminary analysis of the data, 
we decided to consider their responses in the analysis.

The third and last part of the survey consisted of only one assessment (offender
\#40), which the participant made right before exiting the survey and was
analogous to those in the anchoring setting. Through this assessment, we
wished to test any meaningful differences in the predictions of the participants
that had been assigned to the two different treatments. We did not find any
substantial difference in their responses and consequently we omitted the
discussion of these results from the paper.
% should explain what the purpose of this question was

The survey also included three questionnaires, 
which we call {\it perception questions}, 
in which participants were asked to reflect 
and elaborate on the predictions that they had made. 
The questionnaires were located at the end of the first part of the survey
(where feedback was given, after offender $\# 14$),
in the middle (after offender $\# 27$)
and at the end (after offender $\# 39$) 
of the second part of the survey. 
Perception questions, 
which include the participant's self-reported level of confidence, 
accuracy, trust, and use of the algorithmic tool, 
are described in~\textsection\ref{sec:app_perception_questions}. 
For each set of questions, participants were asked 
to refer only to those predictions that they had made 
in the corresponding part of the survey.\footnote{Throughout 
the paper, we will eventually omit the answers
given to the questions in the second questionnaire
% for the following two reasons:
% First, this omission simplifies the presentation of results.
because a technical issue affected the answers given 
by approximately the first 200 participants 
to the second set of perception questions. 
We provide more details about the issue in~\textsection\ref{sec:app_perception_questions}.} 
Participants' demographics were collected 
at the end of the survey (see~\textsection\ref{sec:app_demo}).
We deployed the survey through an interactive web application,
using the \texttt{shiny}~\cite{shiny} and \texttt{shinyjs}~\cite{shinyjs} packages in \texttt{R}.

\subsubsection{Compensation} %  
The payment scheme consisted of a base amount of 
\$1.5 awarded at the completion of the survey 
and of a bonus up to \$5 proportional 
to the predictive performance achieved by the participant in the prediction task. 
The bonus was based on an incentive-compatible payment mechanism 
based on the answers provided in the second part of the survey. 
For the non-anchoring setting, 
only the predictions made by the participant 
with the assistance of the RAI were taken into account. % perhaps I should explain (and double check) why it's incentive compatible:
% inc comp of Brier score is straightforward from the derivation of the first order condition of the expected payoff
% inc comp of accuracy is straightforward
% incentive compatibility of the entire mechanism *should* follow
The computation of the bonus worked as follows.
The total reward was split evenly across the 25 assessments, i.e., the highest possible reward for each assessment was \$0.20. 
Then, for 12 out of the 25 total assessments, 
performance was measured as the accuracy 
of the binary predictions ($Q^b$). 
For the remaining cases,
the bonus was computed according to the Brier score,
a proper score function which has been used by prior work \cite{green2019disparate, jung2020limits}. 
%Therefore the reward for a question 
%in the second part of the survey was equal 
%to $0.20\% \times \sum_{i=15}^{40} [r_i (1-|Y_i-Q_i^b|) + %(1-P^b)(1-(Y_i-Q_i^p)^2)]$
%$0.20\$ \times [P^b (1-|Y_i-Q_i^b|) + (1-P^b)(1-(Y_i-Q_i^p)^2)]$, 
%where $Y_i$, $Q_i^b$, $Q^p_i$ were respectively outcome and answers 
%relative to the $i^{th}$ offender, 
%and $P^b$ was a Bernoulli random variable with mean $12/25$.
% in retrospect an easier payment scheme would have been fine

\subsubsection{Attention checks and exclusion criteria} % 

We designed three ``attention checks'' to assess 
whether participants were reading carefully the offenders' descriptions.
% whether participants were paying adequate attention. 
The attention checks
%in every way 
looked like other vignettes,
except that participants 
were explicitly told what answer to provide for $Q^b$ in the text. 
Specifically, the statement 
``The offender was/was not rearrested in the three years after release'' 
was inserted in the offender's description 
in some random position 
between two other sentences,
with the inclusion of `not' chosen randomly.
Participants who answered incorrectly were
not allowed to proceed with the successive assessments or to retry the survey. 
We placed the first attention check in the first part of the survey 
and the other two in the second part.\footnote{The
second attention check was added only after 
% around 
30 participants had already completed the survey. 
This additional check decreased the likelihood 
that participants would pass all attention checks
by random chance from 1/4(=1/$2^2$) to 1/8(=1/$2^3$).}

To further ensure that participants included in the final analysis 
had spent sufficient time on each question, 
we recorded the time taken to complete 
each offender's assessment and the entire survey as well.
Participants were not informed 
that time was recorded independently of the Mechanical Turk system. 
To calculate the time elapsed per question, 
we created a new timestamp 
when a new offender's profile appeared on the participant's screen 
and another once both predictions were made. 
The difference between the two timestamps 
represents a measure of the time that
the participant spends reading the offender's description 
and making the two predictions. 
Since participants could take breaks,
this measure likely represents an overestimation
of the time that they were fully engaged with the task. 
Participants that spent less than one second on one or more 
of the prediction tasks were excluded from our analysis.  
% Before conducting the experiment,
% we decided to exclude participants 
% that would have spent less than 1 second on one of the predictions.

\subsection{Procedure}\label{sec:procedure} We now provide a brief overview of
how a hypothetical participant navigated our survey
(see~\textsection\ref{sec:app_surveystructure} for more details). Upon accepting
the task on Mechanical Turk and logging into the initial web page of our survey, the
participant was shown a consent form. In case of consent, their identifier was
collected and we checked whether the participant had not attempted or completed
the survey before. If that was not the case, a short description of the task
with a sample question was then displayed. Once an answer was provided, the
participant was shown the full set of instructions. To ensure that participants
paid sufficient attention, information regarding the content of the task (e.g.,
sample of offenders, RAI) and the requirements (e.g., presence of attention
checks) were presented in two consecutive web pages. These two pages could still be
accessed throughout the rest of the survey. Participants were told that the RAI
had been trained to predict re-arrest on the same population of offenders using
a superset of the information available to them and that it was well calibrated,
a property that was explained in detail and illustrated in a plot. Participants
were informed that their compensation would be based on the accuracy of both
their likelihood estimates and binary predictions ($Q^p$ and $Q^b$). Those participants
that were assigned to the non-anchoring setting were not told that only the
predictions they had made after having access to the RAI's recommendation would
be taken into account. This choice was made to ensure that participants invested
an equal amount of effort in all answers and, at the same time, to guarantee
that payments would be similar across settings if having access to the RAI's
predictions led to a large increase in performance. At the end of the
instructions, the participant was shown one offender's description at a time and
was required to make both likelihood estimates ($Q^p$) and binary ($Q^b$) predictions
before proceeding to the following offender. Navigating backward to revise
submitted answers was not possible.

\subsection{Recruitment and participants}\label{sec:recruitment}

The study was advertised on Amazon Mechanical Turk as follows: ``This is a
research study about the prediction of criminal behavior. Participants can
receive up to \$6.5 based upon their own performance. Average total reward is
expected to be \$5''. We limited our pool of participants to workers that had an
historical HIT approval rate higher than 90\%, over 500 Human Intelligence
Tasks (HITs) approved, were located in the US, and had not completed or
attempted the survey before.
%We conducted the study at different times of the day and week during May and
%September 2020. 
The maximum time allowed to submit the HIT on Mechanical Turk was set at 90
minutes. A total of 1438 Mechanical Turk workers tried the survey. Approximately
900 of them failed one of the attention checks and were not allowed to retry the
survey. Less than 10 workers were rejected after the completion of the survey
because they completed one or more prediction tasks in under one second. 
% in retrospect we should have given some partial payment to those workers that
% made it to the second or third attention checks
We were left with valid data from 531 participants. On average, these
participants completed the survey in 28 minutes (standard deviation=13.5,
median=25.4) and earned a bonus of \$3.37 (sd=\$0.47, median=\$3.37). The
average reward was \$12.7 per hour (sd=\$5.9, median=\$11.4).

Participants were mostly male (proportion=62.5\%, total=332), White (76.1\%,
404), and had a college degree (80.2\%, 426). The average age was 38.7 years
(sd=11.9, median=36). College graduates and males were overrepresented in our
sample compared to their prevalence in the US population. Additional details and
comparison with demographics from the Census are presented in
Table~\ref{tab:demo} of \textsection\ref{sec:additional_data}. 
% Approximately
% half of the participants (50.3\%, 267) self reported familiarity with the use of
% algorithmic tools in experiments or in the real world. 

\subsection{Data analysis}\label{sec:data_analysis} Throughout the paper, we use
the following methods and notational shortcuts (inside parentheses).
Correlations on both ordinal and cardinal data are computed using Spearman's
rank correlation coefficient ($S$). We use Kruskall-Wallis test ($KW$) as the
omnibus test of association between a numeric outcome and a categorical factor
variable.  The Kruskall-Wallis test is a
rank-based nonparametric analog to one-way ANOVA. We use Mann-Whitney U test
($MW$) for post-hoc comparisons or comparisons between only two groups.
% check if we are using Wilcoxon somwhere
To test the statistical significance of pairs of differences in means between
independent samples, we use Welch's t-test ($TT$). Standard deviation (sd) and
standard error (se) are sometimes reported together with (or in place of) the
results of these tests. The reported confidence intervals generally rely on the
asymptotic normality of the distribution of the corresponding test statistic.
Confidence intervals for the AUC, however, are obtained using the bootstrap.   
Before conducting the experiment, we chose the significance level of
$\alpha=0.001$ which we use for testing all the hypotheses. Accordingly, we
report only whether the p-values ($p$), which we adjust via Bonferroni correction in
case of multiple testing, are lower or higher than the chosen significance level
($\alpha$). 
%and the paired t-test ($PTT$) in case of independent and dependent samples
%respectively. - for some reason I ended up not using the paired t test.
%Although it should not a big issue given that we are only losing in terms of
%power and I don't think that any of the p-values was low enough to become
%significant, I should go back and check
Throughout the analysis, we make the simplifying assumption that all pairs of
predictions (i.e., likelihood estimates and binary predictions $(Q^p, Q^b)$), both
between and within participants, are independent. 
%This assumption may not hold
%if there was some form of dependence across answers. 
We relax this assumption only in case of
the pre-registered and corresponding revised predictions in the non-anchoring
setting. 

%\subsubsection{Measures of trust and reliance} 
To quantify participants' reliance on the RAI in the second part of the survey
(i.e., offenders \#15-39), we employ metrics based on $Q_P$ (when available),
$Q_{P+RAI}$, and $Q_{RAI}$ for both $Q^p$ and $Q^b$. These measures include
analogs to measures that were adopted by prior work, such as {\it
deviation} \cite{poursabzi2021manipulating}, {\it
influence}~\cite{green2019disparate, green2019principles,
poursabzi2021manipulating}, {\it agreement fraction}, and {\it switch
fraction} \cite{yin2019understanding}. 
% \footnote{Note that the agreement and switch fractions are defined at
% the participant's level, e.g., the agreement fraction for a certain participant
% is the share of assessments for which their final binary predictions match the
% model's. In our work, we generally report summary statistics computed on the
% overall sample.}
Here, we formally introduce only influence. This metric, which is also known as
``weight of advice'' \cite{gino2007effects}, quantifies the magnitude of the
revision of the participant's risk estimate relatively to the difference between
the RAI's recommendation and the participant's initial prediction. In
mathematical terms, for each offender's assessment made by the participant, influence is
defined as $(Q^p_{P+RAI}-Q^p_{P})/(Q^p_{RAI}-Q^p_{P})$. We excluded cases for which
$|Q_{RAI}^p-Q_P^p|\le  5\%$. According to this definition, influence can only be
measured for the predictions made by the participants that were assigned to the
non-anchoring setting. It is 1 if the participant's revised prediction matches
the RAI's, 0 in case of no revision.

\section{Results}\label{sec:results}

Together, the 531 participants made 28015 pairs of predictions ($Q^p$, $Q^b$) on
3521 different offenders. 
%The resulting distribution over the settings and conditions participants were
%assigned to is shown in Table~\ref{ref:random}. 
Approximately half of the participants were assigned to the anchoring setting
(proportion=49\%, total=260) . 
%Around 35\% and 25\% of the participants were assigned to the decreasing and
%controlled conditions, respectively (totals 188 and 130, respectively).
The average time spent on each offender's assessment was 15 seconds (sd=34.2,
median=10). The total time taken to complete the evaluation of the 40 offenders
was 14.9 minutes (sd=8.3, median=12.9) for the participants assigned to the
non-anchoring setting and 11.4 minutes (sd=6.6, median=10.0) for those in the
anchoring setting. The time gap is due to the 25 additional pairs of questions
asked in the non-anchoring setting.\\
In the sample shown to the participants, 42\% of the offenders were rearrested.
The RAI produced an average probability of re-arrest of 42\% (median=40\%) and
predicted re-arrest for 29.5\% of the offenders. The risk estimates made by
participants were slightly higher but of similar magnitude to the RAI's
(mean=50.4\%, median=50\%). But participants predicted that 59.2\% of the
offenders would be rearrested, twice as many as the RAI had predicted.

We now turn to the presentation of our key results. The structure of the rest of the
section is as follows. In~\textsection\ref{sec:mapping_prob_bin}, we discuss how
participants' risk estimates ($Q^p$) did not uniformly map onto binary
predictions ($Q^b$). In~\textsection\ref{sec:study_anc}, we show that
participants did not anchor on the RAI's predictions.
In~\textsection\ref{sec:time}, we show that performance was not related to time
spent on the survey and total time spent on the survey is only a poor proxy for
time truly spent on each question. Last, in~\textsection\ref{sec:pred_vs_dec} we
compare predictions of re-arrest and judicial decisions regarding
incarceration.\footnote{For the interested reader, in the supplementary material
we include the following four analyses and related findings.
In~\textsection\ref{sec:performance}, we show that participants' predictive
performance, both with and without the assistance of the RAI, was lower than the
RAI's with respect to all metrics but the false negative rate.
In~\textsection\ref{sec:performance_self}, we show that, as in
\citet{green2019disparate}, participants' self-reported predictions accuracy was
correlated  with their real accuracy only when outcome feedback was provided.
In~\textsection\ref{sec:race}, we show that, in contrast with past
work~\cite{green2019disparate, green2019principles}, we did not find evidence of
disparate-interactions in the uptake of RAI's predictions.
In~\textsection\ref{sec:order_effect} we show that the ordering of the
offenders' profiles did not affect self-reported trust and reliance on the tool.
Lastly, in~\textsection\ref{sec:app_no_anc}, we examine how participants
assigned to the non-anchoring setting updated their risk estimates and binary
predictions. This subsection extends the discussion
in~\textsection\ref{sec:mapping_prob_bin}.}

%%%%%%%%%%%%%%%%%%%%%%%%%%%%%%%%%%%%%%%%%%%%%%%%%%%%%%%%%%%%%%%%
%%%%%%%%%%%%%%%%%%%%%%%%%%%%%%%%%%%%%%%%%%%%%%%%%%%%%%%%%%%%%%%%
\begin{figure}[t]
\centering
\begin{subfigure}{.5\textwidth}
  \centering
  \includegraphics[width=\linewidth]{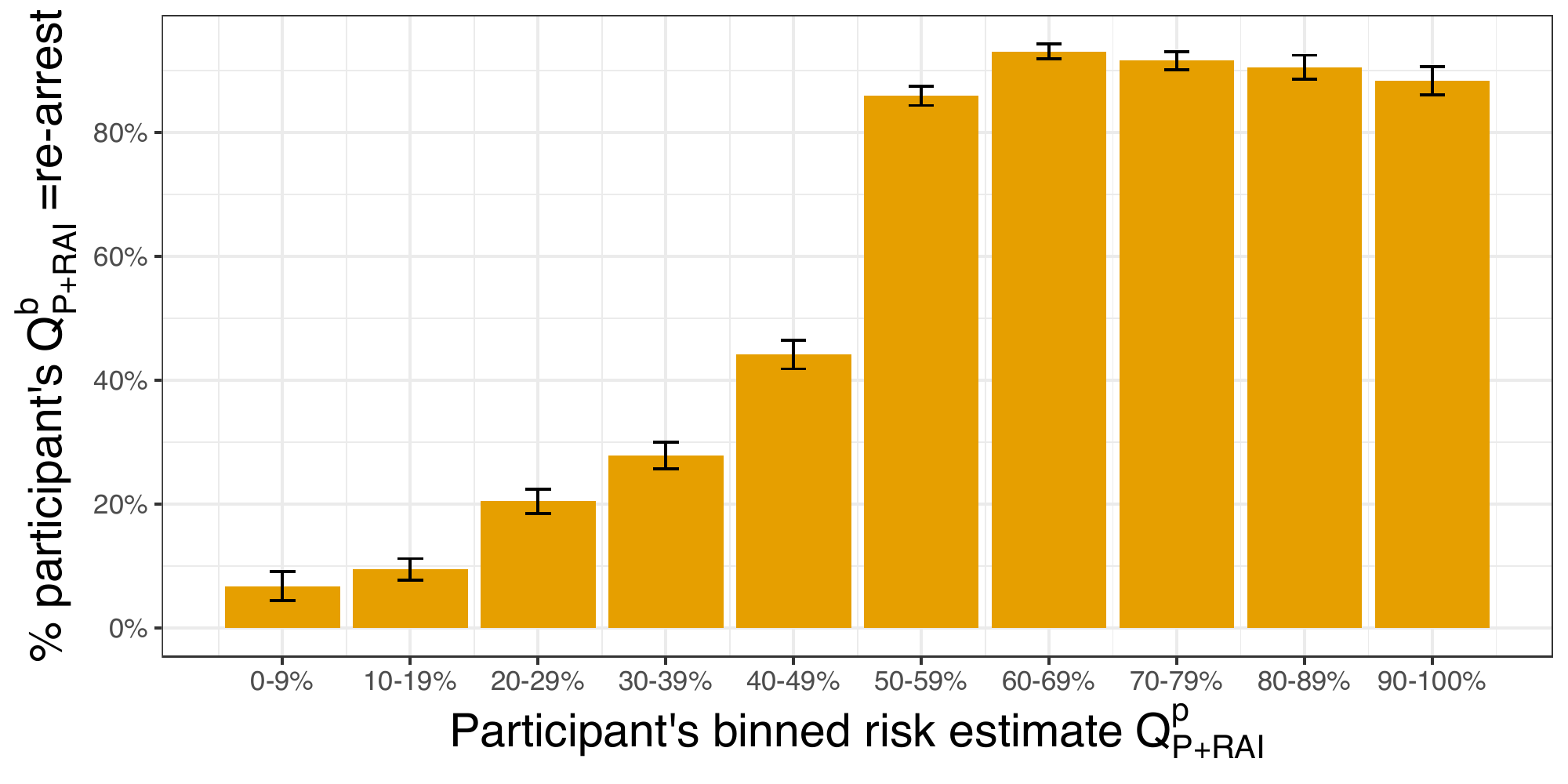}
  %\caption{}
\end{subfigure}%
\begin{subfigure}{.5\textwidth}
  \centering
  \includegraphics[width=\linewidth]{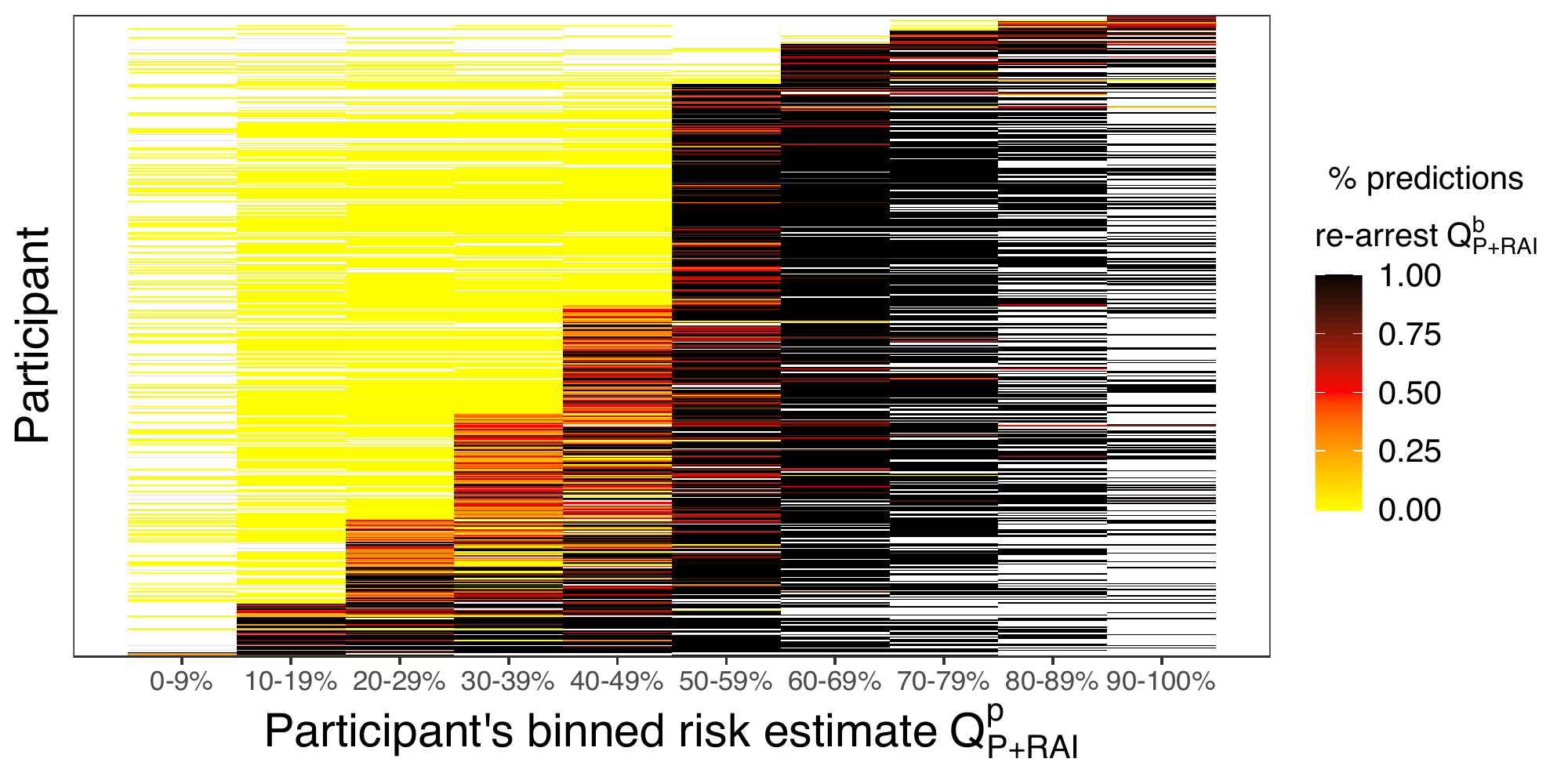}
 % \caption{}
\end{subfigure}%
\caption{Comparison of risk estimates $Q_{P+RAI}^p$ and corresponding binary predictions $Q_{P+RAI}^b$
made by participants in presence of the RAI. %, in the second part of the survey. 
(left) Share of assessments with predictions of re-arrest as a function of the
(binned) risk estimates. Error bars represent 95\% confidence intervals. For
example, participants predicted that the offender would be rearrested in
more than 40\% of the assessments in which they estimated the probability of
re-arrest to be between 40\% and 49\%. (right) Share of assessments with
predictions of re-arrest as a function of the risk estimate and the
participant. Each row corresponds to a different participant. A large fraction
of the participants often---but not always---predicted re-arrests while
assigning likelihoods well below 50\%. Note that participants rarely applied a
consistent probability threshold in predicting
re-arrest.}\label{fig:mapping_probs}
\end{figure}

% \subsection{The same risk estimate does not always map onto the same binary
% prediction}
\subsection{Divergence between risk estimates and binary predictions}
\label{sec:mapping_prob_bin} % old title: Risk estimates and binary predictions

In~\textsection\ref{sec:survey_design}, we described how the payment scheme in
the second part of the survey was designed to incentivize participants to report
their most accurate likelihood estimates and binary predictions. If participants
adopted the profit-maximizing strategy, they would convert their likelihood
estimates $(Q_{P+RAI}^p)$ into binary predictions $(Q_{P+RAI}^b)$ by predicting
re-arrest for all those offenders for whom the value of $Q_{P+RAI}^p$ was 50\%
or greater. Only a small fraction of the participants showed this behavior.
\footnote{Here, we consider only the predictions made by participants in the second
part of the survey and in the presence of the RAI.
We found very similar patterns also for the answers given by
participants in absence of the RAI (i.e., $Q_{P}^p$ and $Q^b_P$) throughout the
entire survey. This indicates that the presence of the RAI did not have any
substantial impact on how participants translate risk estimates into binary
predictions.} Figure~\ref{fig:mapping_probs} shows that participants assisted by
the RAI predicted re-arrest even when their risk estimates were well below the
50\% threshold. Approximately one fourth of the offenders (26\%, se=0.5\%) whose
estimated probability of re-arrest was below 50\% had predictions of
re-arrest. Conversely, one tenth of the offenders (9.1\%, se=0.4\%) whose
estimated risk was larger than 50\% had predictions of no re-arrest. The left
panel of Figure~\ref{fig:mapping_probs} describes this phenomenon by displaying
the share of predicted re-arrests (i.e., mean of $Q_{P+RAI}^b$) as a function
of the corresponding estimated likelihood of re-arrest ($Q_{P+RAI}^p$),
appropriately binned. We observe a gradual increment in the share of predicted
re-arrests as the risk estimates increase. As one could expect, the share of
predicted re-arrests drastically increases around 50\%: A binary prediction is
twice as likely to correspond to re-arrest if the likelihood is just above the
optimal threshold (mean $Q_{P+RAI}^b$=44.1\% for $Q_{P+RAI}^p\in [40\%,49\%]$,
76.4\% for $Q_{P+RAI}^p=50\%$, 90.8\% for $Q_{P+RAI}^p\in[51\%,60\%]$; all
$p_{TT}<\alpha/3$).

The right panel of Figure~\ref{fig:mapping_probs} shows the significant
heterogeneity in the strategies adopted by the participants. Approximately one
third of them (32.4\%, 172 out of 531) always predicted re-arrest when their
probability estimate was greater than 50\% and no re-arrest when their estimate
was less than 50\%. Approximately half of the participants (56.1\%, 298)
followed this strategy in at least 90\% of their predictions. Interestingly,
Figure~\ref{fig:mapping_probs} shows that participants did not even use fixed
thresholds, i.e., for most participants there was no given threshold $t$ for
which $Q^b=$rearrest whenever $Q^p \ge t$.

\paragraph{Consequences for evaluations of participants' predictive performance}
As discussed in \textsection\ref{sec:related_work}, the results of
\citet{jung2020limits} were obtained by soliciting only likelihood estimates
 and converting them into binary predictions by applying a uniform
threshold (50\%) across all participants. Our findings indicate that this
assumed correspondence is not borne out in practice, and analyses based on this
assumption can lead to incorrect conclusions. For example, consider the question
of assessing the predictive performance of our study participants' binary
predictions with converted binary predictions (obtained by thresholding $Q^p$ at
50\%) compared to actual binary predictions $Q^b$.\footnote{For this analysis,
we excluded all risk estimates exactly equal to 50\%. As before, we used
only the risk estimates and predictions made by participants in presence of the RAI.} While
the overall accuracy of the converted predictions is similar to that of the
actual binary predictions ($59.2$\% and $57.6$\% resp., $p_{TT}>\alpha$), other
performance metrics are quite different. The false positive rate is 8.4\% lower
for the converted binary predictions compared to the actual binary predictions
(40.8\% vs. 49.2\% , $p_{TT}<\alpha$), and, instead, the false negative rate is
8\% higher (40.4\% vs. 32\%, $p_{TT}<\alpha$).

\begin{figure}[t]
\centering
\begin{subfigure}{.5\textwidth}
  \centering
  \includegraphics[width=\linewidth]{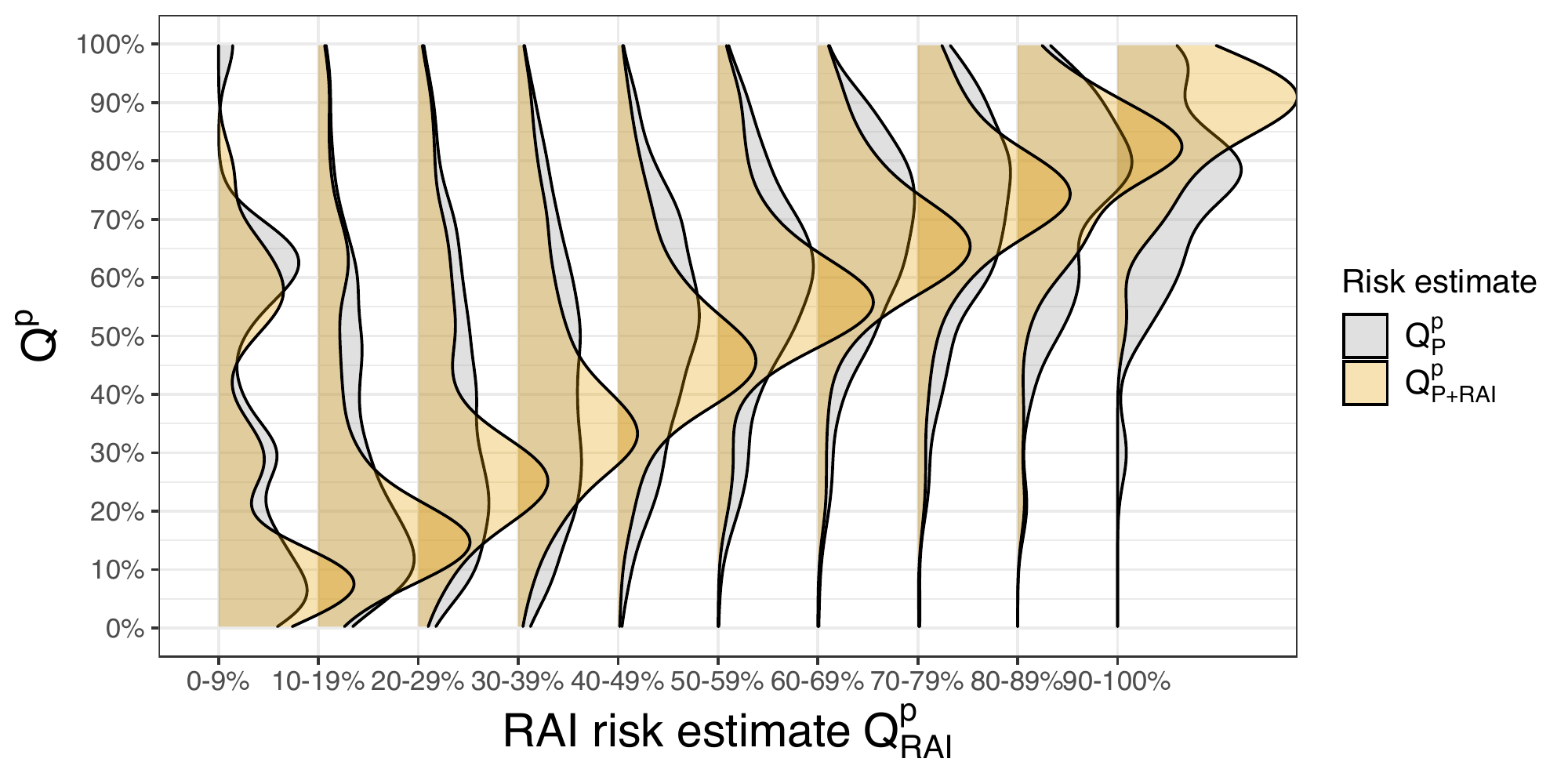}
  %\includegraphics[width=\linewidth]{Figures/Scores_no_anchoring.pdf}
  %\caption{}
\end{subfigure}%
\begin{subfigure}{.5\textwidth}
  \centering
  \includegraphics[width=\linewidth]{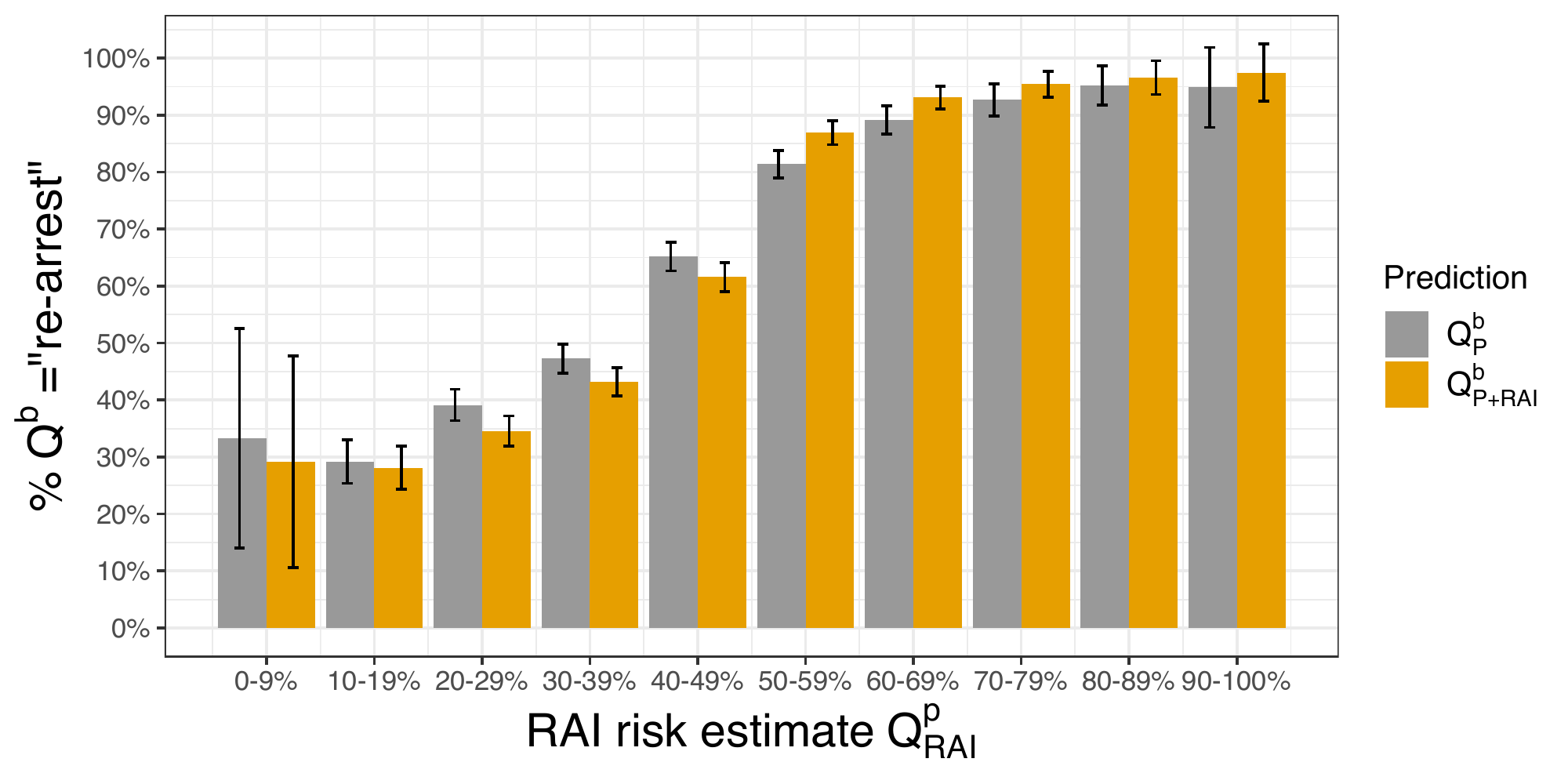}
 % \caption{}
\end{subfigure}%
\caption{Analysis of predictions made by participants assigned to the
non-anchoring setting in the second part of the survey. The two figures show the
distribution of the pre-registered $Q_P$ and revised $Q_{P+RAI}$ predictions
made by participants (vertical axis), grouped by the RAI's risk estimate
$Q^p_{RAI}$ binned (horizontal axis). (left) Density estimates of the
participants' risk estimates $Q^p$. Compared to the pre-registered predictions,
the densities of the revised risk estimates put more mass on the interval
where also the RAI's predictions lie. This indicates that, unsurprisingly,
the participants' revised estimates were closer to the RAI's.  
(right) Share of assessments with predictions of re-arrest in participants'
pre-registered $Q^b_{P}$ and revised $Q^b_{P+RAI}$ binary predictions. Error
bars represent 95\% confidence intervals. The agreement between the RAI's and
the participants' predictions increased after the RAI's prediction was shown,
but the disagreement remained high in some of the buckets, even for low values of the RAI's risk
estimates.}\label{fig:no_anchoring}
\end{figure}

\paragraph{Consequences for evaluations of participants' reliance on the RAI} 
We now examine how participants assigned to the non-anchoring setting updated
their pre-registered probability and binary predictions after they were shown
the RAI. When it came to risk estimates,
Figure~\ref{fig:no_anchoring} shows that participants tended to update their
estimates in the direction of the RAI's recommendation. These revisions were
often substantial (mean influence=37\%, see also
Figure~\ref{fig:analysis_change_influence}; in addition mean of $|Q^p_{RAI}-Q^p_P|=18.8\%$
vs. $|Q^p_{RAI}-Q^p_{P+RAI}|=13.3\%$). 
Participants changed their pre-registered binary answers in 10.2\% (se=0.3\%) of
all predictions, switching in slightly more than half of these cases from a
prediction of re-arrest to a prediction of no re-arrest (see breakdown by RAI's
score in Figure~\ref{fig:no_anchoring}).
%(share of predictions that switch from re-arrest to no re-arrest: $5.7\%$).
However, the overall agreement with the RAI's (binary) predictions only
increased by less than 4\% from the pre-registered to the revised answers (mean
$Q^b_P=Q^b_{RAI}$:$61.7\%$, mean $Q^b_{P+RAI}=Q^b_{RAI}$:$65.4\%$,
$p_{TT}<\alpha$) because participants' final prediction often did not match the
RAI's even if their pre-registered prediction did. It is certainly possible
that, had participants been provided also with the RAI's binary predictions, we
would not have observed such a phenomenon (c.f. ``result 2'' in
\citet{grgic2019human}). We also asked participants to self report for what
share of the assessments they had revised their own predictions after taking
into account the RAI's recommendation. Interestingly, they indicated that they had revised
their risk estimates and binary predictions in 59\% and 34\% of the assessments
respectively (medians=60\% and 20\%). Thus, evaluations based solely on one of
the two types of predictions could lead to drastically different conclusions
regarding participants' (especially self-reported) reliance on the RAI.

\subsection{Absence of anchoring effects}\label{sec:study_anc}
%\subsection{Absence of anchoring effect

In the introduction, we hypothesized that the participants could anchor on the
RAI's predictions. If this had happened, then the risk estimates of the
participants assigned to the anchoring setting would have been closer to the
RAI's than those made by the participants assigned to the non-anchoring setting.
As we might expect, we found that the pre-registered risk estimates of the
participants in the non-anchoring setting were 16\% further from the RAI's than
those of the participants in the anchoring setting. Yet, very surprisingly, the
revised risk estimates made in the non-anchoring setting were 17\% {\it closer}
to the RAI's than those made in the anchoring setting (mean and median absolute
diff. $|Q^p_{P+RAI}-Q^p_{RAI}|$ in the non-anchoring setting: 13.3\% and 8\%
resp., in the anchoring setting=16.1\% and 11\% resp.; $p_{MW}<\alpha$). In
contrast, participants' binary predictions matched the RAI's at exactly the same
rate across the two settings (mean $Q^b_{RAI}\neq Q^b_{P+RAI}:34.6\%$ in both).
We also found that performance did not differ across the two settings: accuracy
(acc. in anchoring=57.4\% and in non-anchoring=57.6\%), false positive rate,
false negative rate, positive predicted values, and the AUC were not
significantly different ($p_{TT}>\alpha$ for all comparisons).\footnote{We conducted analogous analyses after dropping the
predictions that were made in less than 3, 5, and 10 seconds. Again, we found no evidence of anchoring effects.}

In the perception questions, participants self reported similar levels of trust
in the tool across the two settings (76.5\% and 79.3\% of participants in
anchoring and non-anchoring resp.; $p_{TT}>\alpha$), and also of confidence and
accuracy of their own predictions. Participants reported revising their binary
predictions after accounting for the RAI's at approximately the same rate in the
two settings (mean reported revision=31.8\% in anchoring and 33.7\% in
non-anchoring; $p_{MW}>\alpha$). However, the story is different in case of risk
estimates. Participants assigned the non-anchoring setting self reported that
they had revised their answers on average 65\% more often than those in the
anchoring setting after looking at the RAI's recommendation (mean reported
revision=58.9\% in non anchoring, 35.7\% in anchoring; $p_{MW}<\alpha$). Despite
the risk estimates of the participants assigned to the non-anchoring setting
were only slightly closer to the RAI's, their perceived use of the tool to
adjust risk was substantially higher. It seems possible that, by pre-registering
their risk estimates, participants became more aware of the influence of the RAI
on their risk estimates.

To summarise, we did not find evidence of anchoring effects. Instead, we found
the opposite: The predictions made by participants in the non-anchoring setting
were closer (in absolute distance) to the RAI's. These participants also self-reported that the tool had influenced more heavily their risk estimates.

\begin{figure}[t]
\centering
\begin{subfigure}{.5\textwidth}
  \centering
  \includegraphics[width=\linewidth]{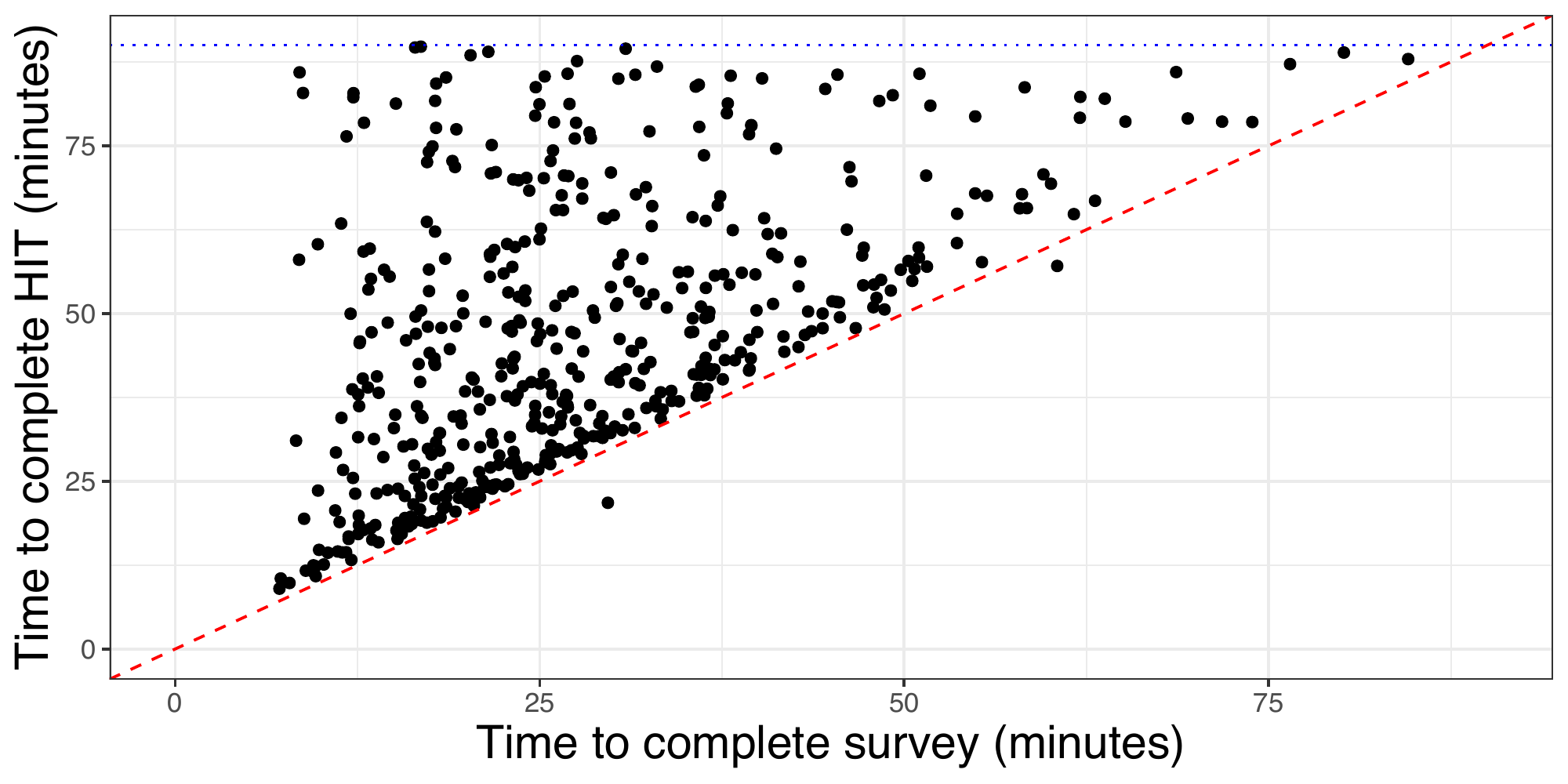}
\end{subfigure}%
\begin{subfigure}{.5\textwidth}
  \centering
  \includegraphics[width=\linewidth]{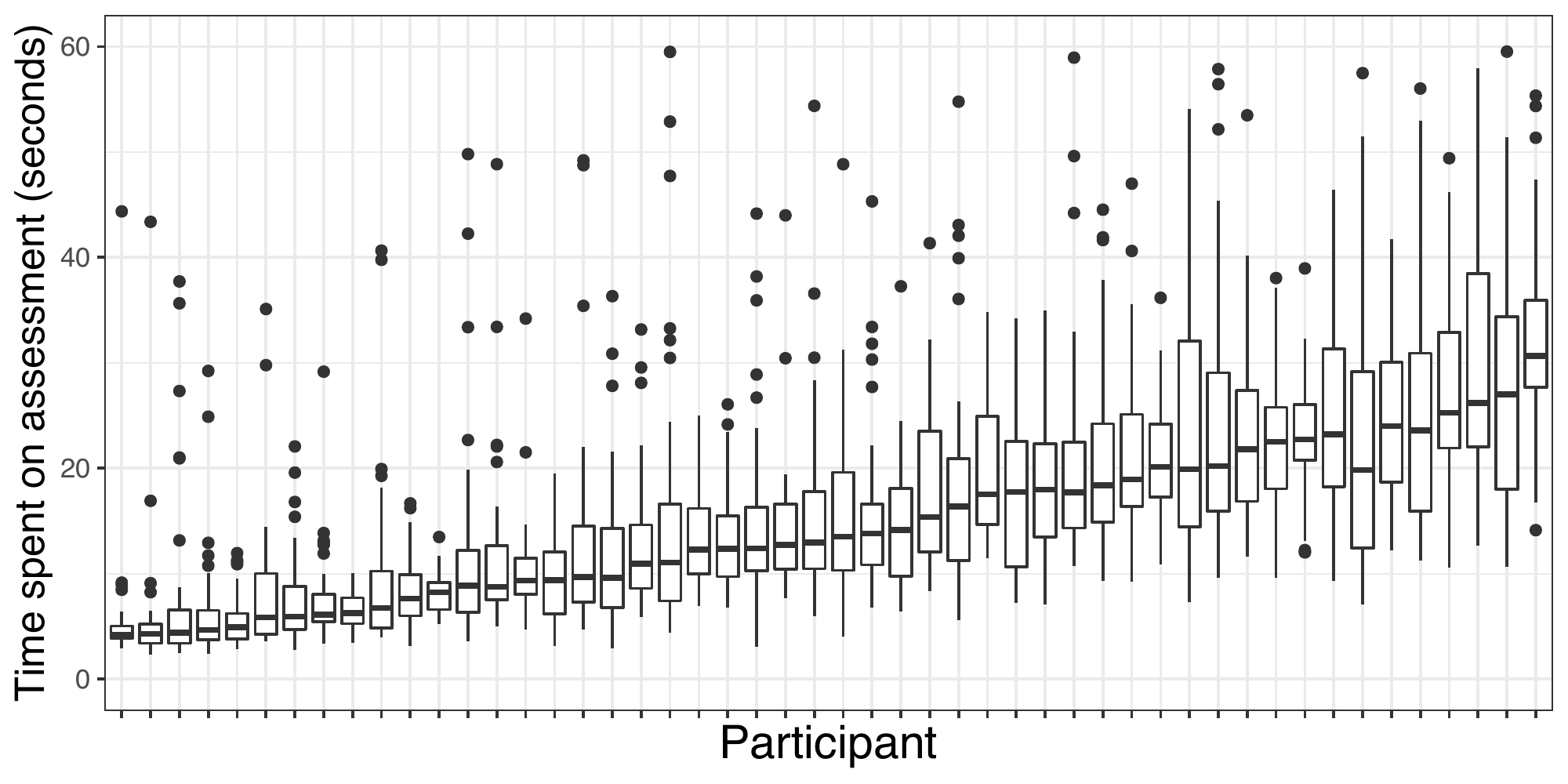}
 % \caption{}
\end{subfigure}%
\caption{Analysis of time. (left)  Time taken to complete the HIT on Mechanical
Turk (vertical axis) as a function of the time spent on the survey (horizontal
axis) for each participant. The blue dotted line corresponds to 90 minutes,
which was the maximum time allowed on Mechanical Turk for the submission of the
HIT. Each black dot corresponds to an individual participant. The observations below
the diagonal line correspond to participants that started the survey before
accepting the HIT. If the time on Mechanical Turk and on the survey were similar
for each of the participants, the observations would lie close to the dashed red line
at 45 degrees.  We observe that the time taken to complete the HIT was generally
substantially larger than the time spent on the survey. (right) Boxplot of the
time spent on each offender's assessment by every participant. The whiskers
extend from the hinges to the smallest or largest values at most 1.5$\cdot$IQR
of the hinge. For visualization purposes we randomly sampled 100 participants
from those that had been assigned to the anchoring setting and limited the
length of the vertical axis.}\label{fig:analysis_time}
\end{figure}

\subsection{Time spent on the survey as a poor proxy for time spent on the
assessment}\label{sec:time}
%Time spent on the survey is a poor proxy for the time spent on the assessment
%of each offender}

Participants completed all the steps in the survey, from login to submission, in
an average time of 28 minutes (sd=13.5). In comparison, the average time
recorded on Mechanical Turk, from the acceptance to the submission of the HIT,
was 68\% greater, averaging 46.7 minutes (sd=20.2). The left panel of
Figure \ref{fig:analysis_time} shows the time spent on the survey and on
Mechanical Turk for each participant. The time spent on the survey was on
average 89\% greater than the time recorded on the platform. For 27.9\% of the
participants (139 out of 499)\footnote{We could not match all survey
participants to their Mechanical Turk identifiers.} the submission of the HIT
took 5 minutes less than the completion of the survey; for 43.5\% of them
(217) it took 10 minutes less, and for 63.1\% of them (315) it took 20 minutes less.
% I did a mistake and recorded only the time spent on (1) reading the payment
% and (2) answering the first sample question -- not talking about this
Figure \ref{fig:analysis_time} also reveals that the time spent on the survey
 varied considerably across participants, from less than 10 minutes up to more than
one hour. An analysis at a more granular level reveals that the assessment of
an individual vignette was carried out on average in 15 seconds (sd=34.2,
median=10). Once we excluded the pre-registered and revised predictions made by
the participants assigned to the non-anchoring setting in the second part of the
survey,\footnote{These participants made two pairs of predictions for each
offender and the revised prediction was made more quickly than the
pre-registered one (mean time of revised=9.4 seconds, sd=21.7).} the average
time was $17.5$ seconds (sd=39.2, median=12.2).

The substantial difference between the mean and the median raises the following
question: Is the time taken to complete the survey (divided by the number of
questions) a reliable proxy for the time spent on each question? Equivalently,
do participants spend the same amount of time on all question?\footnote{If this
was the case, then we could infer---with some certainty---bounds for the time
taken to complete an individual assessment based on the total time spent on the
survey. In this analysis we only considered all 40 predictions made by the
participants that were assigned to the anchoring setting.} The right panel of
Figure~\ref{fig:analysis_time} shows that participants generally did not
allocate their time equally across questions. On average, they spent more than
one tenth of their time on the assessment of only one offender (out of 40, mean
proportion of time per participant=12.1\% and median=7.7\%) and one fifth of
their time on four assessments (20.3\%, median=16.6\%). Once the four
assessments that took longest were removed for each participant, the time spent
that each of them spent on the assessments decreased on average by 19.4\%. We
also found, as might be expected, that participants tended to spend more time on the first question of
each part of the survey. For example, in the first part of the survey,
participants spent on average almost 9 seconds more on the first question than
in the others (average time first question=26.8 and median=17.6 seconds, others
mean=17.9 and median=12.4; $p_{MW}<\alpha$). Similarly, the first question in
the second part of the survey took twice as much time as the others (average
time first question=30.6 seconds, others=17.2; $p_{MW}<\alpha$). We could not
identify any other clear reasons for why some predictions took longer
than others. It is likely that participants simply took breaks from the survey.
Nonetheless, they allocated the time proportionally to the number of questions
in each of the two parts of the survey (i.e., on average 37.8\% of their time on
the first part of the survey that contained 35\%=14/40 of the offenders).

We also examined whether time was related to predictive performance. For the
participants that had been assigned to the anchoring setting, the time spent on
the entire survey was weakly correlated with the accuracy of the binary
predictions ($\rho_{S}=0.3$, $p<\alpha$). However, we observe that this
correlation is mainly due to the participants that completed the surveys very
quickly and also achieved very low accuracy. Assessments on which predictions
were accurate did not take longer than inaccurate predictions on average
($p_{TT}>\alpha$) and only the median time was slightly higher (median time
accurate=12.4 seconds and inaccurate=11.9; $p_{MW}<\alpha)$. Given the large
variance in the time spent on the assessments across participants, we also
tested a similar hypothesis. For each participant, we ranked their 40
predictions in increasing order of time. The mean ranking of the predictions
that were accurate was, perhaps unexpectedly, {\it lower} than that of the those that
were inaccurate.  One possible explanation is that some vignettes are easy to get right, and thus do not require much time to complete.  

In summary, we found that the time taken to complete the HIT on Mechanical Turk
was a severe overestimation of the time that participants spent on the survey.
The time spent of the survey was also a poor proxy for the time spent on each
vignette, given that participants typically spent a substantial share
of their time on only a few assessments. Lastly, we found little evidence of any
association between time spent by the participants on the assessment and
accuracy of their predictions.

%%%%%%%%%%%%%%%%%%%%%%%%%%%%%%%%%%%%%%%%%%%%%%%%%%%%%%%%%%%%%%%%
%%%%%%%%%%%%%%%%%%%%%%%%%%%%%%%%%%%%%%%%%%%%%%%%%%%%%%%%%%%%%%%%

\subsection{Predictions are not decisions}\label{sec:pred_vs_dec} In our study,
participants were asked to provide likelihood estimates and binary predictions of
offenders' re-arrest outcomes. As we discuss in this section, our findings concerning
how humans update their predictions when presented with the RAI's
recommendations---and findings from these types of studies more generally---do
not readily translate into implications for decision-making. Firstly, there is
the clear issue that our study participants are not judges, and are not being
presented with the full set of information that judges would have access to in
real world decision-making settings. However, we would not be surprised if
conducting our experiment on a population of judges produced largely the same
qualitative findings regarding risk \textit{predictions}. The bigger issue that
we wish to draw attention to here is that risk is just one of many factors that
judges are asked to consider in their \textit{decisions}.  

In the context of sentencing, considerations of risk are often secondary to factors such as the severity of the offense and
restitution to victims. The offender's risk of re-arrest often enters into decisions indirectly through considerations of prior criminal history, which is a leading predictor of future recidivism.   %Indeed,
%judges decide whether the offender should be incarcerated based on a series of
%factors, among which is the offender's risk of re-arrest.
%But also punishment (i.e., retributive justice), restitution to victims, and
%threat to the public matter.
For instance, the Pennsylvania Commission on Sentencing has adopted a set of
sentencing guidelines that ``provide sanctions proportionate to the severity of
the crime and the severity of the offender’s prior conviction
record''~\cite{pacodemanual}.
% I should cite the legal document but should find out how to cite it:
% http://www.pacodeandbulletin.gov/Display/pacode?file=/secure/pacode/data/204/chapter303/s303.11.html&d=
% basic matrix:
% http://pcs.la.psu.edu/guidelines/sentencing/sentencing-guidelines-and-implementation-manuals/7th-edition-amendment-5-1-1-2020/303.16-a-basic-sentencing-matrix/view
These guidelines are primarily a function of two factors: the offense gravity
score (OGS), which is a measure of the gravity of the most serious crime among
the offender's current charges; and the prior record score (PRS), which is a
composite measure summarizing prior criminal records. Higher values of OGS and
PRS correspond to more serious offenses and more severe prior criminal
histories \cite{hyatt2016use}.
% The guidelines identify five sentencing levels that, in turn, correspond to a
% certain ranges of sentencing options available to courts. For example, the
% only option in the lowest level are restorative sanctions, such as community
% service. In the highest level, the most lenient option is 12 months of
% confinement.
While the recommendations are advisory, judges need to provide a written
justification when their sentencing decisions deviate from the recommended
range. 

\begin{figure}[t]
\centering
 \includegraphics[width=0.8\linewidth]{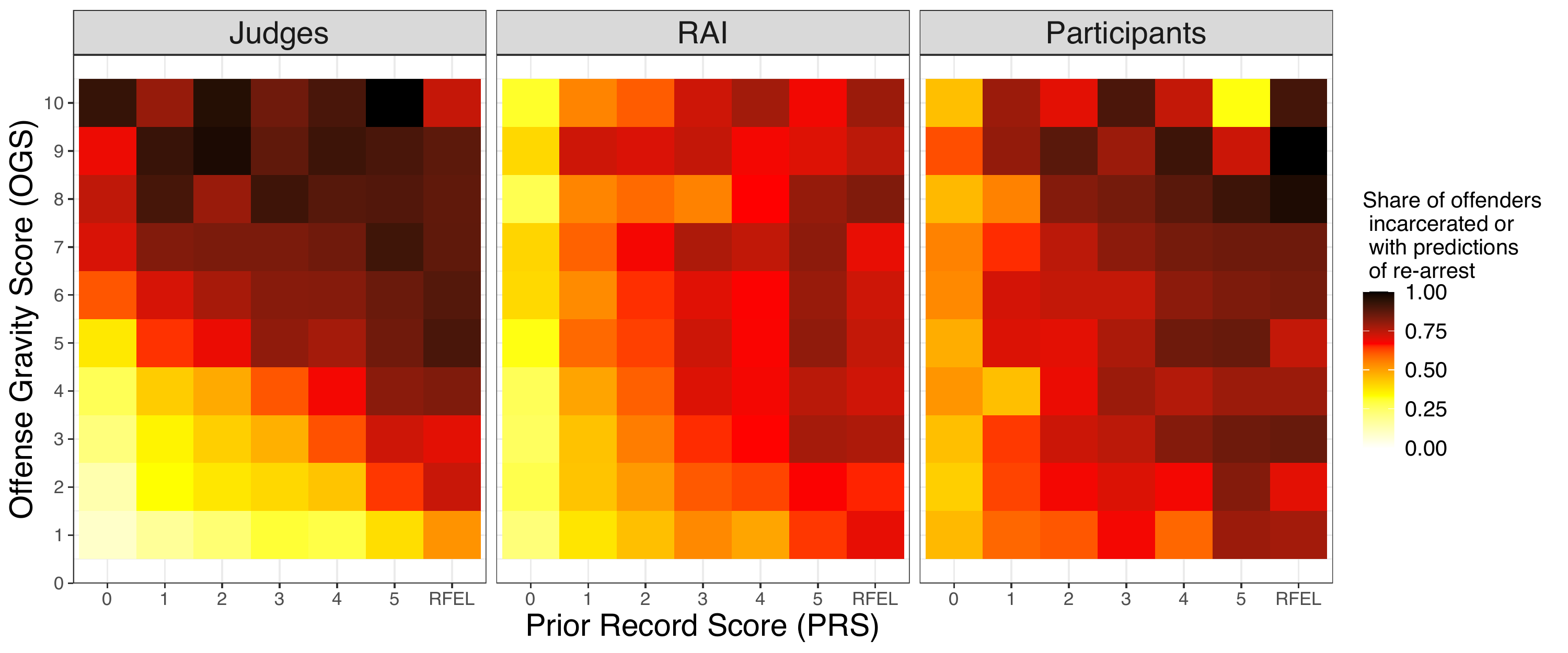}
\caption{Comparison of judicial decisions of incarceration, binary predictions
of the RAI, and binary predictions of participants (both $Q_{P}$ and
$Q_{P+RAI}$). The threshold for converting the RAI's risk estimates into binary
predictions for this analysis was set such that the overall share of predicted
re-arrests was equal to the incarceration rate. The color in each cell is
proportional to the share of offenders that were incarcerated (judge) or with
predictions of re-arrest (RAI and participants), for specific values of the
offense gravity score (OGS) and prior record score (PRS). In this figure, the
heatmaps for judicial decisions and RAI's predictions were generated using all
offenders in the test set.}\label{fig:pred_vs_dec}
\end{figure}

Our data contain information not only on re-arrest outcomes, but also on
judicial sentencing decisions.  
This allows us to compare judicial decisions of whether to sentence the offender
to a period of incarceration to the predictions of
re-arrest made by participants and RAI. Figure \ref{fig:pred_vs_dec} shows judicial decisions along with
participants' and RAI's predictions broken down by levels of the OGS and the
PRS. For judges, we observe that the likelihood of incarceration
increases with both the PRS and OGS. But the pattern is more complex in case of
participants' and RAI's predictions of re-arrest: The likelihood of a predicted
re-arrest appears to increase with the PRS, but it does not seem to be associated with
the OGS.  

We investigated how the likelihood of incarceration and the predictions of
re-arrest made by participants and RAI depended on the PRS and the OGS by
fitting three separate logistic regression models of the form $outcome\sim PRS +
OGS$. We treated both scores as numeric (repeated felony offender, ``RFEL'', was
converted into a score of 6) and set the threshold for the RAI such that the
share of predicted positives was equal to the share of offenders that were
incarcerated. The coefficients of the OGS were $0.05$, $0.08$, and $0.49$ for
the models targeting predicted re-arrest by participants, by the RAI, and
incarceration respectively (all $p<\alpha$). These results would indicate that,
ceteris paribus, an increase in the OGS by one level was associated to a 63\%
increase in the odds ratio of the likelihood of incarceration but only to a 5\%
increase in that of the participants' predictions of re-arrest (see the full
results in Table~\ref{tab:log_models}).
% exp(0.5)=0.6 The coefficients of the PRS and the interaction term were similar
%across the models.

To take a concrete example, consider the subgroup of offenders charged with drug
offenses. Among these offenders, 68\% of those charged with felonies were
incarcerated, compared to only 21\% in case of misdemeanors. Yet, the rate of
re-arrest was virtually identical across the two types of charges (47\% and 45\%
respectively).\footnote{Even if we assumed that imprisonment helped reduce
re-arrest rates (and evidence suggests that this is unlikely to be the case
\cite{nagin2013deterrence}), the large gap in incarceration rates could still not be
explained by differences in the likelihood of re-arrest.} In comparison,
participants and the RAI, once equalized for overall rates, predicted re-arrest for only a slightly
higher share of the felony offenders (share of predictions of re-arrest for
felonies: RAI=43\% and participants=66\%; misdemeanors: RAI=52\% and
participants=59\%). This indicates that there are cases where differences in
decisions are unrelated to the risk of recidivism (here, the decision of whether
to incarcerate). 

Because risk predictions are not aligned with the OGS, and the OGS is one of the
primary determinants of criminal sentences, we should not expect findings of how
judges' predictions change when provided with RAI information to be directly
informative of likely changes in the resulting decisions.  
Even if our findings that participants revised their risk predictions when
presented with the RAI's recommendation generalized to judges, it is unclear what
effect, if any, those revised assessments would have on sentencing outcomes.

% if imprisonment was indeed criminogenic, then the "paradox" would be even
% greater

\section{General discussion}

\subsection{Limitations}\label{sec:limitations}
There are several limitations related to 
the generalizability of our results. 
First, despite our strict exclusion criteria, 
the analysis in~\textsection\ref{sec:time}
revealed that 
a substantial share of the 
assessments 
%considered in our analysis 
were carried out very quickly, 
often in less than 5 seconds.
% included 
This indicates that participants often did not fully read the offenders'
descriptions or did not think carefully about their answers. The post-hoc
analyses, however, have revealed that our conclusions were largely unaffected by the
presence of these predictions in the data. It is also unclear whether our
findings would generalize to other experimental setups (such as those with
different incentive structures), populations of participants, domains, or to
real-world decision-making settings. For example, we might expect judges to
exhibit many of the cognitive biases exhibited by our survey participants
\cite{guthrie2000inside} and suspect they might also overestimate their own
predictive abilities \cite{demichele2018intuitive}, which may result in lesser
reliance on the RAI. Consequently, the impact of the RAI on their judgment may be
different than what has been observed in our experiment. Another potential
limitation is that our results could have been affected by our explanations of
the incentive structure, information regarding the RAI, and framing of the
questions. We also note that our pool was composed of Amazon Mechanical Turk
workers that were $20$--$40$ years old, a demographic group that is likely 
% extremely 
tech-savvy and more open to the introduction of technologies than other
populations~\cite{redmiles2019well}. As such, it is not a representative sample
of the broader US population, and is not demographically representative of the US judiciary. Lastly, it is possible that, due to the negative
experiences that workers often have on Mechanical Turk \cite{fort2011amazon,
mason2012conducting}, running the experiment on an alternative platform could
lead to different results, especially those discussed in \textsection\ref{sec:time}.

\subsection{Discussion}
In our study, participants often predicted re-arrest %in binary feedback %would happen
even when assigning a low numeric score to the likelihood of that event.
% that the future occurrence of a re-arrest even when they deemed its likelihood to be low. 
% Why did we observe such a behavior?
% 
% We identified two reasons that could explain this behavior.
% 
We offer four hypotheses that could explain this behavior.
The first is poor numeracy skills: 
% Research has shown that 
Even highly educated individuals tend to struggle with 
simple numeracy questions~\cite{black1995perceptions, schwartz1997role, lipkus2001general, visschers2009probability}. 
% It is possible that 
Some of our participants may have not realized 
that the expected profit-maximizing strategy is to predict that the given offender would be re-arrested if and only if they believed
the risk of rearrest is greater than 50\%.
A second explanation might be that participants’ subjective likelihood judgments did not simply reflect the estimated frequencies of events \cite{manski2004measuring, evans2004if}, e.g., due to an asymmetry in the perceived cost of errors.
This hypothesis seems plausible, even though participants were incentivized through the payment scheme to report calibrated probability estimates. 
%A second explanation might be the asymmetry in the perceived cost of errors. %When making binary predictions, it is possiblethat participants confused predictions with decisions (of whether to incarcerate or not). If this was the case, their choices could % reflect preferences on the inherent trade-off reflect cost sensitivity between limiting threat to public safety or the offender's freedom.
%For example, participants an asymmetry in the perceived cost of errors might. errors in the perceived cost of errors. 
%However, this hypothesis is not supported by our finding that, conditional on the defendant's risk, the association between the participants' prediction of re-arrest and the gravity of the offense committed by the offender was weak (\textsection\ref{sec:pred_vs_dec}). 
A third explanation might be lack of care: 
% is suggested by the observation that,
% despite characterizing all predictions, 
% is that 
We found that, while the observed phenomenon held in general,
% 
% this phenomenon was more prevalent 
it was more common among the predictions that took the least time. Consequently,
it is possible that some of the participants made the predictions quickly and
without paying adequate attention, e.g., satisficing. The fourth
explanation might be that participants did not understand the slider instrument,
despite being provided with a clear set of instructions (see
\textsection\ref{sec:app_partial_instructions}). However, given that this
instrument is prevalent in current practice \cite{roster2015exploring} and it
generally has high concurrent validity with respect to other types of
instruments such as Likert scales and radio buttons \cite{couper2008designing},
this hypothesis seems unlikely to explain our results. Future work might explore
why this phenomenon occurred. 

It is of paramount importance to employ measures
that are comprehensible to the human and can also be effectively and
consistently estimated by the human. It is also equally important that the RAI's
recommendations be easily understood by the human. Different ways to communicate
the RAI's prediction can help decision makers calibrate their trust in the
RAI~\cite{zhang2020effect, lai2019human} and, potentially, also reduce
variability in their decisions. The comparison between our results around how
people often switch their (binary) predictions not in the direction of the RAI's
and the findings in~\citet{grgic2019human} likely represents an example of the
importance of such design choices. Another notable implication of this result is that
the type of prediction on which participants' predictive performance and
reliance on the RAI's recommendations are evaluated can affect the conclusions
that researchers draw from the study. In our experiment, for example, an
analysis that focused only on binary predictions would have neglected the role of
the RAI in influencing participants' risk estimates. While the choice of the
type of prediction might be context-dependent, the potential limitations of the
assessment should be recognized and acknowledged.

One notable and unexpected finding in our work
is the absence of anchoring effects.
We designed the second part of the survey 
to determine how much (if at all) 
participants anchored on the RAI's predictions.
For binary predictions, we found no difference 
in the agreement between the participants 
and the RAI across the two settings.
For risk estimates, not only did our participants 
not anchor on the RAI---participants 
who were shown the RAI directly
gave predictions further from the RAI's than
participants that had pre-registered their estimates 
% without the RAI's assistance.
before seeing the RAI's outputs.
One possible explanation of this behavior 
is that participants felt that they 
had to---or wanted to---demonstrate a sense of agency. 
According to this hypothesis, 
participants assigned to the non-anchoring setting, 
who had already delivered risk estimates 
without the assistance of the RAI,
% would have
would have been more comfortable with accepting the RAI's predictions,
having already demonstrated their agency.
% often defaulted to the model's estimate, as we did observe. 
% Instead,
At the same time,
participants assigned to the anchoring setting 
might have felt compelled to offer 
predictions that differed from those of the RAI
to demonstrate that they were performing the task
earnestly and not simply copying the RAI.
% may have  to make predictions that differed from the RAI's. 
Another possible explanation 
for this phenomenon might be 
that the users who pre-registered their predictions also
% participants
had the opportunity to notice 
% felt the
how similar the RAI's predictions were to theirs
% similar to theirs 
and might consequently have trusted the RAI more.
% started relying on the RAI more. 
However, these users 
% did not translate into (even) higher levels of self-reported trust in the tool.
did not report higher levels of trust in the tool.
Yet they reported making a heavier use of the tool's predictions. 
This finding opens an interesting 
new direction for future research: 
If these results held in other experiments,
could we improve experts' perceived and actual reliance on algorithmic tools
% (particularly in contexts where the tools is much more accurate than the human) 
by eliciting predictions from the experts 
both before and after revealing the RAI's recommendation?

Like past works, %our analysis has also found that,
we found that even when participants were shown the RAI's predictions, 
they continued to underperform the RAI in terms of predictive accuracy.
% Analogous conclusions have been drawn in case of predictions of loan defaults
%Similar results have been seen 
%with predictions of loan default~\cite{green2019principles}.
It seems possible that this observation 
may characterize many forecasting settings 
where both the RAI and humans have low accuracy. 
\citet{tan2018investigating} demonstrated that
even if the predictions made by the RAI 
and the human alone were combined,
the resulting unavoidable error may still be very large. 
We note that in our experimental setting
we would not expect participants' predictions 
to outperform those of the RAI. 
This is because the RAI has access to all of the features 
that participants are able to consider, 
and is trained to optimize for predictive accuracy.  
In practice, judges have access to information that the tool does not. 
Further studies of human-in-the-loop systems 
are needed to examine the influence of the overlap in information sets 
between humans and the RAI on predictive performance, 
participants' trust in the tool, and decision-making.  
% We speculate that other factors influencing the efficacy
% of human-RAI hybrid systems may also depend on whether
% they have access to the same or to complementary information,
% and what sorts of explanations (if any) accompany the RAI predictions.
% and the presentation (e.g., with explanations) of the RAI to the human 
% could improve the complementarity between the two agents. 
It is likely that even in the case of complementary information sets, 
or where participants have access to more information than is available to the tool,
it will be difficult for humans to match the RAI's performance 
without carefully crafted feedback.  

Unsurprisingly, the time that participants actually spent on the survey was
substantially lower than the time taken to submit the HIT and did not even
represent a reliable proxy for the time spent on the assessment of a single
vignette. In addition, participants who spent more time on the survey 
did not, in general, achieve higher predictive accuracy. Ideally, especially in tasks
characterized by high degrees of uncertainty such as the prediction of criminal
recidivism, researchers might want to design compensation structures in which
the assigned reward is proportional to the effort made by the participant. While
time itself is not a perfect measure of effort, future work could employ
exclusion criteria based on the time that participants spend on each assessment.
Time spent is also worth taking into account when 
assessing the generalizability of study findings, especially with respect to
real-world decision-making in high-stakes settings. For example, researchers
could run post-hoc analyses to assess whether their conclusions still hold once
the predictions that participants made quickly are dropped. In our experiment,
as we already mentioned, the main results still held even when these predictions
were not considered in the analysis.
% obviously this is kind of challenging because (i) if participants know about
%this, they will just spend more time It is hard to see how assessments that are
%carried out in less than (say) 5 seconds  would contribute to answering  the
%research question of interest or have any ecological validity for real-world
%high-stakes decision making. 

% recheck this part
Lastly, in~\textsection\ref{sec:pred_vs_dec} we showed that, 
% differently from the case of 
unlike judicial decisions, the predictions made by participants
were weakly associated to the seriousness of the offender's current charge.
For drug felonies and misdemeanors, predictions were %almost
nearly orthogonal to the gravity of the offense.
This finding sheds light on the limits of the use 
of human predictions of criminal recidivism 
as proxies for judicial decision-making:  
% should rephrase sentence recalling the impact of RAIs
Even if studies found large effects of the impact 
of RAI's recommendation on human predictions (which they have not),
further investigation would be required to understand 
how or whether those would translate to decisions.
Such an analysis would not only require understanding 
for which offenders decisions and predictions would diverge,
but also how the introduction of the RAI would 
reshape the existing decision-making framework. %through which the recommendation to the court is made. 
For example, in the Pennsylvania sentencing system, 
the newly adopted RAI is used to determine 
when a pre-sentencing investigation report is to be generated, 
which would provide judges with additional information on the offender~\cite{parai2020}.
% https://sentencing.umn.edu/sites/sentencing.umn.edu/files/pennsylvania_overview_of_the_sentencing_risk_assessment_instrument_2017.pdf
In the New Jersey pretrial system,
the RAI represents a building block of the 
decision-making framework, but it is not its only element~\cite{njdmf2018}. 
There, the RAI is also used to decide 
whether a summons or a warrant should be issued. 
Thus, the RAI potentially affects not one but many sequential decisions 
made at several stages of the pipeline 
by interacting with other elements in a larger framework.
% do not citation bc cannot find the information on the web
% talk about DMF (or perhaps throw it into the related work section

\section{Acknowledgments}
We are grateful to PwC USA for funding this research through the Digital
Transformation and Innovation Center sponsored by PwC. We also deeply thank
Shamindra Shrotriya and Pratik Patil for providing valuable feedback on the
survey.

%%
%% The next two lines define the bibliography style to be used, and
%% the bibliography file.
\bibliographystyle{ACM-Reference-Format}
\bibliography{extracted}
\nocite{*}

%%
%% If your work has an appendix, this is the place to put it.
\newpage
\appendix

\section{Additional details on methodology}\label{sec:app_racial_bias}
\paragraph{Assessing the candidate RAIs for racial predictive bias}
One of the questions we sought to answer with our study is whether the extent or
manner in which participants responded to RAI's predictions showed indications of
racial bias as previously found in \citet{green2019disparate} and 
\citet{green2019principles}.  To ensure that our results would not be confounded by the
use of an RAI that itself systematically over or underestimates the likelihood
of re-arrest for certain racial groups, we performed a group-level calibration
analysis of all four models being considered for use in the experiment.

We first assessed the model's racial calibration properties via logistic
regression following the approach of~\citet{skeem2016risk}. This entails a
comparison of the relative fit of the following three nested logistic regression
models: $Y\sim \text{score}$, $Y\sim \text{score}+\text{race}$ for ``intercept
bias'', and $Y\sim \text{score} +\text{race} + \text{score}\times\text{race}$
for ``slope bias'' through a likelihood ratio test or a Wald test for the
coefficient of the added variable, where $Y$ indicates the re-arrest (Y=1 if
re-arrest occurs, 0 otherwise). None of the models showed intercept bias (all
$p>\alpha$) and only XGBoost presented slope bias (all others $p>\alpha$). Due
to the likely misspecification of the logistic regression model adopted in the
test, we additionally relied on the chi-squared test of
\citet{fogliato2020fairness}. We separated the RAI's predictions by offenders'
racial groups and divided them into bins of width $0.1$ ($[0,0.1),
[0.1,0.2),\dots,[0.9,1]$). The test revealed an overestimation of the risk for
White offenders (compared to Black offenders) in the logistic regression and
XGBoost models for the predictions in the bins $[0.3,0.4)$ and $[0.7,0.8)$
respectively.

We then examined the models for error rate balance. All four models showed false
positive rates on Black offenders that were more than twice as large as those on
White offenders (around 35\% and 13\% respectively), but also lower false negative rates for
Black offenders (around 35\% and 67\% respectively).  This is expected in settings where the
outcome base rates differ across groups, as is the case in the present study.
Positive predictive values were higher for Black offenders (around 66\% vs.
60\%). All differences were statistically significant (all $p<\alpha$). Lastly, we
checked whether the models produced similar AUCs across racial groups, a
criterion named accuracy equity \cite{dieterich2016compas}. The AUC of the
predictions on Black offenders was slightly higher than the AUC of those on Whites across all models (around
$70\%-71\%$ and $68\%-69\%$ respectively). 
%Given the similarity in the performance and properties across the four models, 
%we decided to adopt the Lasso as the RAI in our experiment. 

%%%%%%%%%%%%%%%%%%%%%%%%%%%%%%%%%%
%%%%%%%%%%%%%%%%%%%%%%%%%%%%%%%%%%

%%%%%%%%%%%%%%%%%%%%%%%%%%%%%%%%%%%%%%%%%%%%%%%%%%%%%%%%%%%%%%%%
%%%%%%%%%%%%%%%%%%%%%%%%%%%%%%%%%%%%%%%%%%%%%%%%%%%%%%%%%%%%%%%%
\section{Additional results}

This section is organized as follows. In
\textsection\ref{sec:performance} we compare participants' and RAI's predictive
performances. In particular, we show that participants always underperformed the
RAI. In~\textsection\ref{sec:performance_self} we show that, similarly to the
results of \citet{green2019disparate}, participants' self-reported predictive
accuracy was correlated  with their real accuracy only when outcome feedback was
provided. In \textsection\ref{sec:race} we
show that, in contrast with past work \cite{green2019disparate,
green2019principles}, we did not find evidence of disparate-interactions in the
participants' uptake of the RAI's predictions. However, their risk estimates
overestimated the risk of re-arrest for Whites compared to Blacks and their
binary predictions produced higher false positive rates for Blacks. In
\textsection\ref{sec:order_effect} we show that the order in which the
offenders' profiles were shown did not affect participants' trust and reliance
on the RAI. Lastly, in \textsection\ref{sec:app_no_anc} we extend the results around
the impact of the RAI on participants' predictions presented in
\textsection\ref{sec:mapping_prob_bin}.

\subsection{Evaluation of predictive performance: Participants performed worse than the RAI 
according to all metrics other than the false negative rate}\label{sec:performance}

\begin{figure}[t]
\centering
  \includegraphics[width=\linewidth]{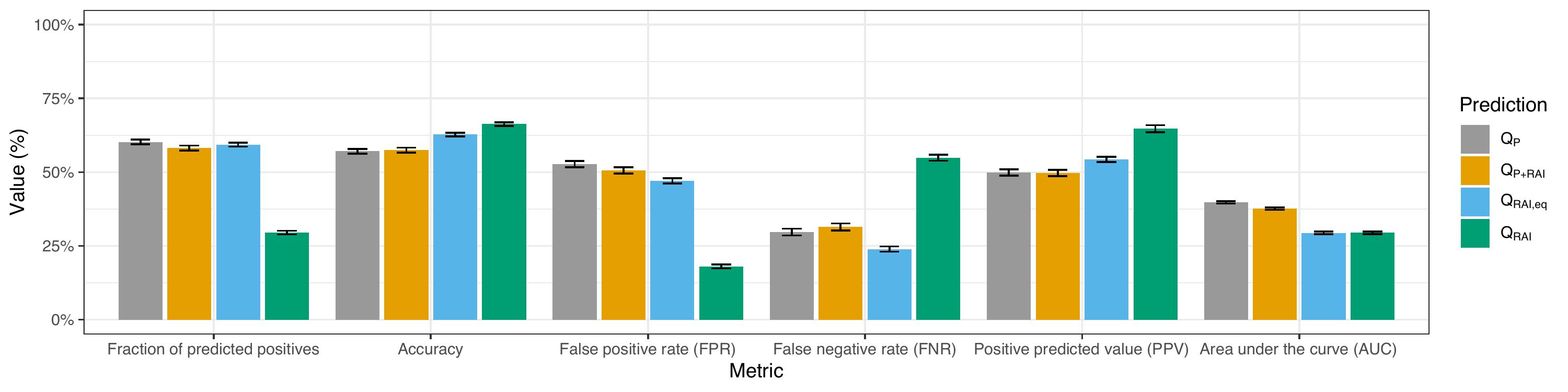}
\caption{Performance metrics relative to the predictions made by participants
alone ($Q_P$), participants assisted by the RAI ($Q_{P+RAI}$), and by the RAI
alone with risk estimates converted into binary predictions using the threshold
of 50\% ($Q_{RAI}$) and another for which the fraction of predicted positives
was equal to the participants' ($Q_{RAI, eq}$). The metrics considered are fraction of
predicted re-arrests, accuracy, false positive rate (FPR), false negative rate
(FNR), positive predicted vaues (PPV), and area under the curve (AUC). The first
five metrics are based on the binary answers given by the participants ($Q^b$),
whereas the AUC is based on the risk estimates ($Q^p$). Error bars indicate 95\%
confidence intervals.}\label{fig:performance_metrics}
\end{figure}

In this section we analyze the predictive performance of the predictions made by
our study participants, both with and without the RAI, across a range of
different metrics. Binary classification metrics such as accuracy and false
positive rates were computed based on participants' responses to the binary
prediction question $Q^b$. Area under the curve calculations are based on
participants' probability predictions $Q^p$. For this analysis, $Q_P$ includes
all answers made by participants in absence of the RAI both in the first and
second parts of the survey. 

One challenge in comparing the binary classification performance of the RAI to
participants' predictions is that the fraction of cases for which the RAI's predicted
probabilities exceed $0.5$ is very different from the fraction of cases in which
participants predicted re-arrest.  
%In~\textsection\ref{sec:mapping_prob_bin}, 
% we showed that performance varied with the type of answer
% that was elicited from the participant. 
% In this section, we measure performance adopting the most natural choice. 
% Fraction of predicted positives, accuracy, false positive rate,
% false negative rate, and positive predicted values 
% are based on the participants' binary predictions ($Q^b$); 
% instead, the area under the curve (AUC) is based on the risk estimates ($Q^p$).
To make the participants' and RAI's predictions comparable, 
we obtained additional binary predictions from the RAI's risk estimates 
using a threshold for which the fraction of predicted positives 
for the RAI was equal to the participants'.  This is denoted by $Q_{RAI, eq}$ in the results.

Figure~\ref{fig:performance_metrics} summarizes our findings.  
We observe substantial differences between the RAI's and the participants' metrics.
On average, the RAI outperformed their predictions 
across all metrics considered other than the false negative rate, 
while the rate-equalized RAI outperformed their predictions 
on all metrics including the false negative rate.   
There is no notable difference in the metrics 
between the predictions that were made by participants alone 
and those in presence of the RAI, thus we only focus on the latter.   
%To simplify the presentation, in the following we will only refer to the latter.

Participant predicted re-arrest in twice as many assessments as the RAI did
(mean for $Q_{P+RAI}$=58.2\% and for $Q_{RAI}=29.5\%$; $p_{TT}<\alpha$).
The accuracy of their predictions was 8.8\% lower than the RAI's (acc. for
$Q_{P+RAI}=57.5\%$ and $Q_{RAI}$=66.3\%; $p_{TT}<\alpha$). The false positive
rate of their predictions was three times higher than the RAI's, but only
slightly higher than that of the rate-equalized RAI (FPR for $Q_{P+RAI}=50.6\%$,
$Q_{RAI, eq}=47\%$, $Q_{RAI}=18\%$; all $p_{TT}<\alpha$). In contrast, the
predictions of the RAI presented higher false negative rates than the
participants' and the RAI with equalized rate's (FNR for $Q_{P+RAI}=31.4\%$,
$Q_{RAI, eq}=23.9\%$, $Q_{RAI}=54.9\%$; all $p_{TT}<\alpha$). The positive
predicted values of the RAI and RAI with equalized rate were 31\% and 9\% higher
than the participants' (PPV for $Q_{P+RAI}=49.7\%$, $Q_{RAI, eq}=54.4\%$,
$Q_{RAI}=64.8\%$; all $p_{TT}<\alpha$). The area under the curve produced by
the RAI's risk estimates was 13\% higher than the participants' (AUC for
$Q_{P+RAI}=62\%$, $Q_{RAI}=70.1\%$). 
%Overall, the RAI with equalized rate performed better than participants across all metrics.
%Except for the false negative rate, the RAI's performance metrics were substantially better than the participants'. 
% here the samples were dependent, I should have used the paired t-test

When we looked at the performance of individual participants 
instead of averaging over all participants,
we did find evidence of human performance, in some cases, 
exceeding that of the RAI.  % also analyzed what fraction of participants achieved an accuracy higher than the RAI's on their observed batch of offenders. 
In the second part of the survey, 
31\% of the participants had an accuracy higher than the RAI's.
Thus, while most participants were unable to improve upon or match 
the predictive performance of the RAI 
even when presented with the RAI predictions, some did. 
%Despite being shown its predictions, participants were unable to use the information to improve on its predictions.

Past work has noted that participants' accuracy, in presence of feedback, slowly
increased with the number of examples that were shown \cite{jung2020limits}. In
our study, participants received feedback on only the first 14 assessments
(compared to 50 in past work), but they could still learn by seeing the RAI's
predictions. We did not find evidence of any increase in accuracy throughout the
survey. The accuracy of participants remained stable across the assessments in the
first part of the survey (accuracy first 7 offenders=57.7\%, second 7
offenders=57.3\%) and was not higher for the pre-registered predictions.\\

\subsection{Evaluation of predictive performance: Participants' self-reported predictive
accuracy was correlated with actual accuracy only in presence of
feedback}\label{sec:performance_self} In the perception questions, we assessed
whether participants could correctly guess their own accuracy and the confidence
in their predictions (see all questions in
\textsection\ref{sec:app_perception_questions}). In the question regarding accuracy,
participants were asked what share of their binary predictions they thought were
accurate. In the first part of the survey, where participants were provided with
outcome feedback after each vignette, participants' perceived accuracy was
strongly positively associated to their real accuracy ($\rho_S=0.49$;
$p<\alpha$). When feedback was removed in the second part of the survey,
participants were unable to correctly guess their own accuracy: The
self-reported accuracy was uncorrelated with their actual performance
($\rho_{S}=-0.05$; $p>\alpha$). 
%Participants' guesses were off on average by 11\% (1.59 questions out of 14)
%and 19\% (4.81 questions out of 25) in the first and second parts of the survey
%respectively. 
In the second part of the survey, the average perceived performance was
substantially larger than the real performance (perceived=59.8\% vs. real=57.5\%
in first part, perceived=65.7\% vs. real=57.5\% in second part). 42\% of
participants overestimated their own accuracy in the first part of the survey,
compared to 58\% in the second part.

When asked about confidence in their own predictions, the participants that
reported higher levels of perceived accuracy were also more confident in their
own predictions (in the first part the mean self-reported accuracy of
participants that were confident=66.2\%, neutral=58\%, and not confident=46.2\%;
all $p_{MW}<\alpha$. Similar results were found for the second part). There
was a slight increase in the confidence levels of the participants between the
first and second part of the survey, with just 12.4\% reporting that they were
`not confident' in their predictions in the second part, compared to 19.6\% in
the first part. Overall, we found that throughout the survey participants became
more optimistic about their accuracy and slightly more confident in their own
predictions, potentially due to the absence of feedback.

We also asked participants to evaluate the accuracy of their own predictions
compared to those of other participants. More than half of the participants thought that
their own accuracy was approximately equal to the median accuracy (54.4\% and
55\% in the first and second parts of the survey respectively). A large share of
participants deemed their own performance to be higher than the median accuracy
and this share increased from the first to the second part if the survey (30.3\%
and 37.5\% of participants in first and second parts respectively). When
feedback was given, their evaluations were associated to the real ranking
($p_{MW}>\alpha$ for the comparison of substantially higher than median vs.
around the median, $p_{MW}<\alpha$ for the others). In the second part of the
survey, where feedback was removed, perceived ranking was no longer
significantly associated to real ranking (all $p_{MW}>\alpha$).

\subsection{Racial disparities in predictions}\label{sec:race}
%The potential reduction of existing inequalities across sociodemographic groups
%is a key argument in favor of the deployment of risk assessment instruments.
%Given the deterministic nature of the RAI, disparities can be assessed before
%deployment, up to randomness in the population. In human in-the-loop decision
%making settings, it is often unclear whether the interaction between decision
%makers and RAI would mitigate or exacerbate existing inequalities. 
In this section we investigate whether the predictions that participants made in
presence of the RAI suffered from predictive bias and could lead to disparate
impact. We found that these predictions presented higher false positive rates on
Black offenders, but more severely overestimated the risk for White offenders
compared to Blacks. In contrast to \citet{green2019disparate}, we found no
evidence of racial bias in the uptake of RAI's predictions. As a reminder,
in~\textsection\ref{sec:app_racial_bias} we show that the RAI's predictions
presented higher false positive rates on Black offenders and produced a lower
area under the curve for Whites, but passed the calibration tests. 

Participants largely overestimated the share of re-arrests for both racial
groups (predictions of re-arrests=70\% and 51.8\% for Black and White offenders
respectively). They performed slightly better on the Black population of
offenders in terms of accuracy and area under the curve (accuracy for
Whites=$56.3\%$ and for Blacks=$59.7\%$, $p_{TT}<\alpha$; AUC for
Whites=$60.3\%$ with 95\% confidence interval $[59.3\%,61.3]$ and for
Blacks=$62.4\%$ and 95\% confidence interval $[61.7\%, 63.2\%]$). The other
metrics presented substantial differences across the two racial groups. The
false positive rate on Black offenders was higher by 14\% compared to Whites
(FPR for Whites=46.5\% and for Blacks=60.7\%, $p_{TT}<\alpha$) but the false
negative rate was lower by 18\% (FNR for Whites=39\% and for Blacks=21.3\%,
$p_{TT}<\alpha$). Positive predicted values were 58.3\% for Blacks and as low as
43.5\% for Whites ($p_{TT}<\alpha$). We also conducted two test of calibration
(see~\textsection\ref{sec:app_racial_bias} for more details on these tests).
Probability predictions suffered from both intercept and slope bias in the
calibration test via logistic regression (coef. of White race=-0.45 and
interaction=-0.54 respectively; both $p<\alpha$)~\cite{skeem2016risk}.
Predictions also failed the calibration test via chi-squared ($p<\alpha$ for
all $\lfloor Q_{P+RAI}\cdot 10\rfloor\geq 5$), where the share of re-arrests for
Blacks was higher than for Whites across all bins. 

We checked whether the RAI's influence on the predictions of the participants
depended on the offender's racial group. We tested this hypothesis on two sets
of predictions: (i) the pre-registered and revised risk estimates in the
non-anchoring setting; and (ii) all risk estimates made without and with the
assistance of the RAI.\footnote{Note that only the metric computed in (i) follows the definition of influence presented in \textsection\ref{sec:data_analysis}.} In (ii), we considered only those offenders that had at least
3 predictions of each type, averaged the answers, and then computed the
influence. For both (i) and (ii), we found that the RAI exerted approximately
the same influence across racial groups ($p_{MW}>\alpha$ for both). In contrast
to the findings of~\citet{green2019disparate}, we found that, in cases where the
risk estimates made by participants alone were lower than the RAI's, the
influence of the RAI was {\it not} larger for Black offenders neither in (i) nor
in (ii) ($p_{MW}>\alpha$ for both).

\subsection{The order of the offenders' profiles did not affect participants' interactions with the RAI}\label{sec:order_effect}

As we mentioned in~\textsection\ref{sec:survey_design}, in some cases the
offenders' profiles were not randomly ordered. More specifically, the
participants were assigned to one of three conditions, that we call
\textit{random}, {\it controlled}, and {\it decreasing}. In the `random'
setting, participants were shown a randomly sampled set of offenders. In the
other two conditions, we sampled a set of offenders on which the RAI achieved
the same accuracy as in the population (i.e., 16 or 17 out of 25 predictions
were correct). In the `controlled' setting participants were presented with the
offenders' profiles sorted in random order. In the `decreasing' setting,
participants were first shown the offenders for whom the RAI had made accurate
predictions and then those for whom the RAI's predictions were
inaccurate.\footnote{The randomization scheme differed across the two dimensions
of variation due to the aforementioned technical issue rather than for
methodological reasons. We first collected an initial batch of surveys with only
the random setting. Then, all participants of the second
batch of surveys were mistakenly assigned to the decreasing condition. All
successive participants were then assigned to the controlled condition. 
Since participants were not aware of which of condition they would be assigned
to (the survey and the instructions were always the same), the only other source of
sampling bias likely consisted of the variation in the population of
participants on the crowdsourcing platform when the surveys were published. We
tried to ensure randomization by running the last batches in similar days of
the week and times of the day. We could not find significant differences in the
demographics of our pool of participants and time that they spent on the survey.} See
Table \ref{tab:random} for the distribution of participants across conditions.
Here we only focus on the comparison of the controlled and decreasing
conditions. 

% do we all agree on this title?
We initially expected the participants assigned to the decreasing condition to
exhibit more trust in the RAI and their risk estimates to be more heavily
influenced by the RAI compared to those of the participants assigned to the
controlled condition. We found little or no evidence of such phenomena: Our
results show that, in absence of feedback, participants' trust and reliance on
the RAI were not impacted by the accuracy of its predictions on the initial set of cases. 

None of the performance metrics differed across the two settings other than for the area
under the curve, which was larger by almost 4\% for the predictions made in the
controlled condition (AUC for controlled=62.2\% and 95\% c.i. $[61.1\%,63.4\%]$,
for decreasing=$58.6\%$ and c.i. $[57.3\%, 59.8\%]$; all $p_{TT}>\alpha$ for
other comparisons). The risk estimates made in the two conditions were similarly 
close to the RAI's (mean $|Q^p_{P+RAI}-Q^p_{RAI}|$ for controlled=$16\%$
and for decreasing=$17.5\%$; $p_{MW}>\alpha$). In the last set of perception
questions, participants self reported trusting the RAI at approximately the same
rate across the two conditions (81.5\% and 80.3\% in controlled and decreasing
respectively; $p>\alpha$). Interestingly, trust was not affected by the level of accuracy
of the RAI's predictions: Among the participants assigned to the decreasing condition, the
same share of participants reported trust in the tool in the second and third
sets of perception questions. We also observed no notable
differences in the levels of confidence and self-reported use of the RAI across
the two conditions. 

\begin{table}[t]
\centering
\begin{tabular}{lcc}
  \toprule
  & Anchoring & Non-anchoring \\ 
\midrule
   Random & 20.2\% (107) & 20\% (106) \\ 
   Controlled & 12.1\% (64) & 12.4\% (66) \\ 
   Decreasing & 16.8\% (89) & 18.6\% (99) \\ 
   \bottomrule
\end{tabular}
\caption{Distribution of participants across the 2 settings (anchoring, non-anchoring)
and the 3 conditions (random, controlled, decreasing). 
Raw counts are reported inside parentheses.}\label{tab:random}
\end{table}

\subsection{Participants often revised their risk estimates when provided with
the RAI's prediction, but were less likely to revise their binary predictions in the direction one would expect}\label{sec:app_no_anc}

%\subsection{Participants often revised their risk estimates when provided with the RAI's prediction, but were less likely to revise their binary predictions, and not in the direction one would expect}\label{sec:nonanc}
In this section, we provide a longer discussion on how participants assigned to
the non-anchoring setting revised their predictions after gaining access to the
RAI's. These results extend those in
\textsection\ref{sec:mapping_prob_bin}.
%We now discuss whether and how participants assigned to the non-anchoring setting 
%revised their predictions after gaining access to the RAI's.\footnote{It is possible that the only fact that participants expect the RAI's predictions to become available affects their own predictions~\cite{paravisini2013incentive}.} 
%We found that participants tended to substantially revise their risk estimates,
%but less often their binary predictions 
%and not always in the same direction as the RAI's.

Participants' predictive performance 
did not significantly improve when they were presented with the RAI's predictions (accuracy for $Q_P^b=56.7$ vs $Q^b_{P+RAI}=57.6\%$, $p_{TT}>\alpha$).  
The main exception is in the case of AUC, 
for which we detect a statistically significant improvement of 2.8\% 
(mean AUC for $Q_P^p$=60.1\% and 95\% conf. int. $[59.2\%, 61\%]$, 
mean AUC for $Q^p_{P+RAI}=62.9\%$ and 95\% conf. int. $[62\%, 63.8\%]$). 
However, this does not mean that participants 
did not revise their predictions when provided with the RAI's.
This result should be expected in light of the fact that participants often revised only their risk estimates in the direction of the RAI's.

\begin{figure}[H]
\centering
\begin{subfigure}{.5\textwidth}
  \centering
  \includegraphics[width=\linewidth]{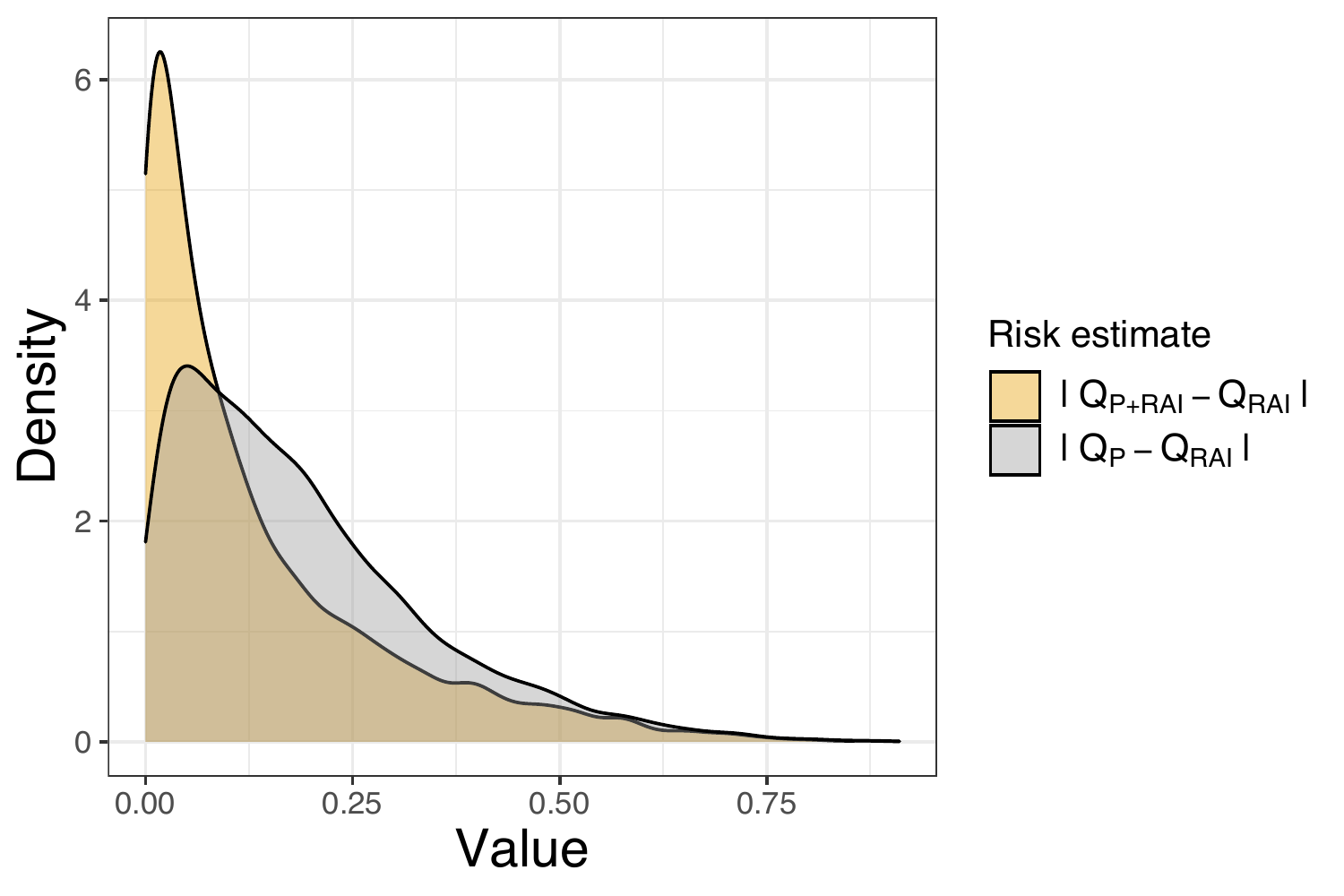}
\end{subfigure}%
\begin{subfigure}{.5\textwidth}
  \centering
  \includegraphics[width=\linewidth]{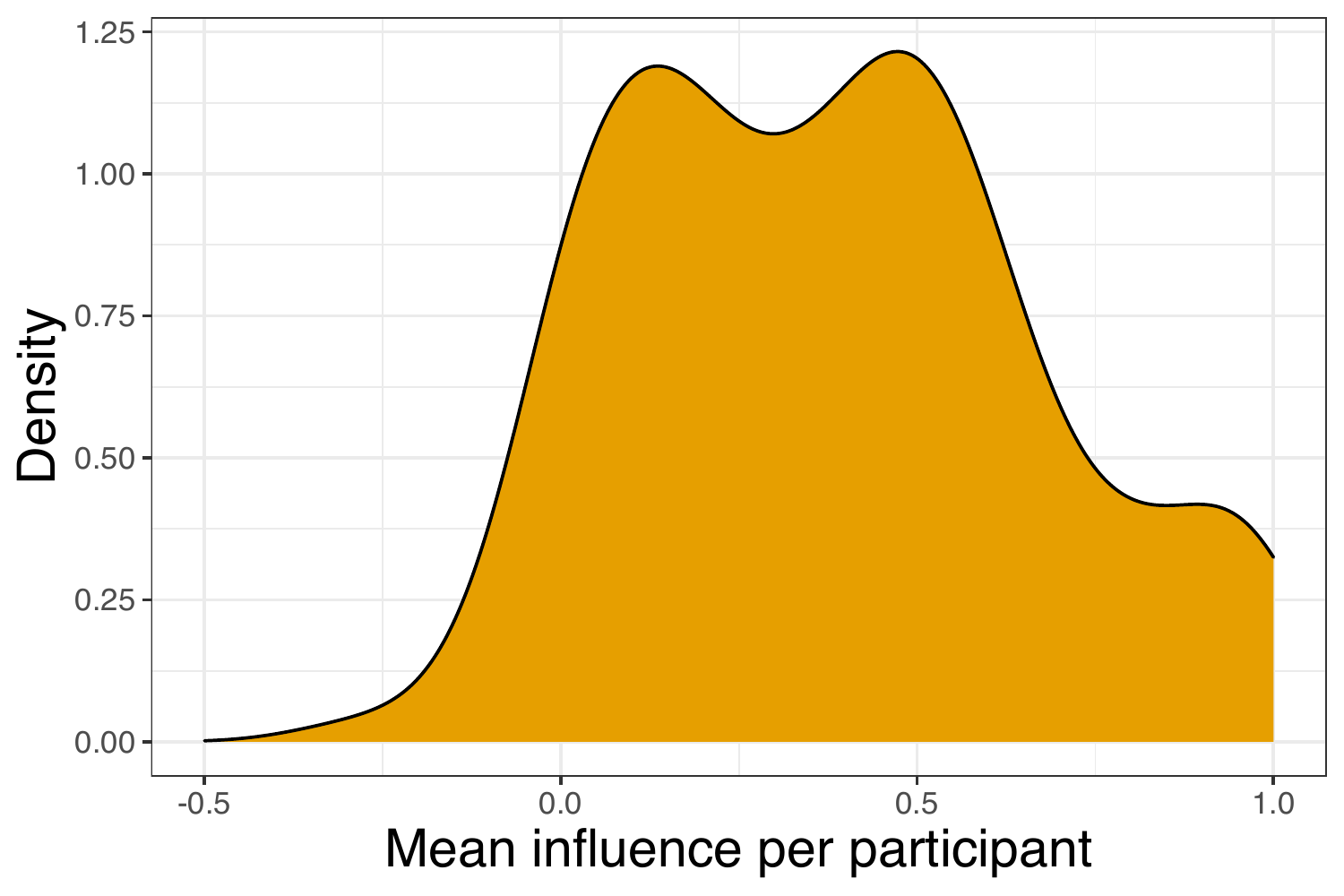}
 % \caption{}
\end{subfigure}%
\caption{Analysis of risk estimates in the second part of the survey for the
participants that were assigned the non-anchoring setting. (left) Kernel density
estimate of the absolute differences of the participants' initial and revised
risk estimates from the RAI's ($|Q_{P}-Q_{RAI}|$ and $|Q_{P+RAI}-Q_{RAI}|$
respectively). The figure shows that participants
 tended to update their estimates in the direction of the RAI's.
(right) Kernel density estimates of the mean influence of the RAI on the risk
estimates by participant. We observe that participants' revised estimates were on average
closer to their initial predictions than to the
RAI's.}\label{fig:analysis_change_influence}
\end{figure}

\begin{figure}[H]
  \begin{subfigure}{.45\textwidth}
  \centering
  \includegraphics[width=\linewidth]{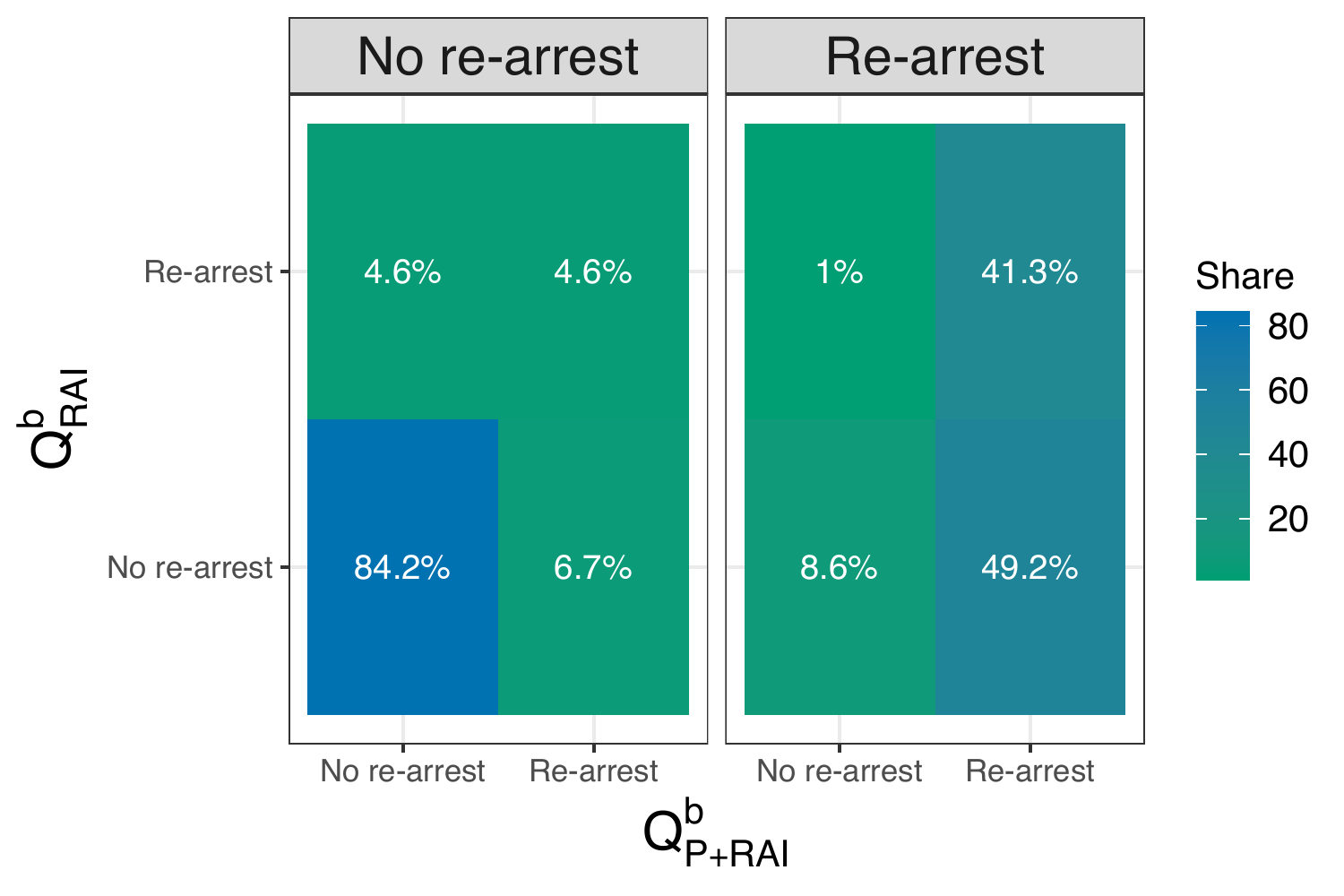}
  \end{subfigure}%
  \begin{subfigure}{.55\textwidth}
  \centering
 \includegraphics[width=\linewidth]{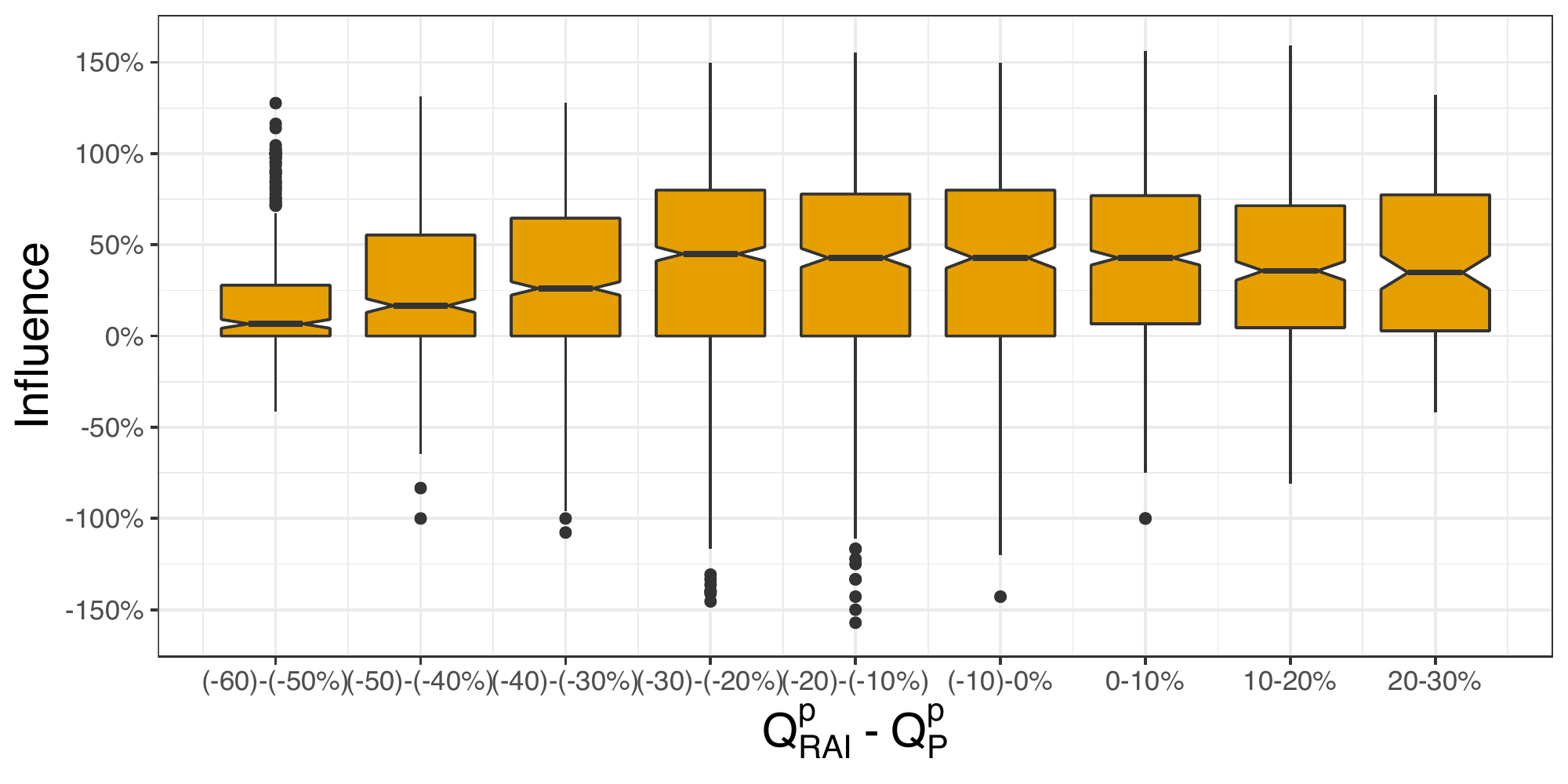}
 % \caption{}
\end{subfigure}%
\caption{Analysis of binary predictions in the non-anchoring setting. (left) The
heatmaps correspond to pre-registered binary predictions corresponding to no
re-arrest (left) and re-arrest (right). The binary predictions of RAI and the
final revised answers of participants lie on the vertical and horizontal axes
respectively. The percentages correspond to the fraction of observations falling
in each category. For example, the percentage in the bottom left corner
indicates the share of all pre-registered predictions corresponding to no
re-arrest (approximately 40\% of all predictions) that matched the RAI's
(binary) prediction and that were not revised. The bottom right corner of the heatmap on the
left and the top left corner of the heatmap on the right represent interesting
cases of participants switched prediction even though their pre-registered
answers matched the RAI's predictions. (right) Boxplot of the influence of the
RAI on participants' risk estimates grouped by the difference between the RAI's
predictions ($Q^p_{RAI}$) and the participants' pre-registered risk estimates
($Q^p_P$). The whiskers extend from the hinges to the smallest or largest values
at most 1.5$\cdot$IQR of the hinge and the notches extend to 1.58$\cdot$IQR
/$\sqrt{n}$. If participants were to blindly rely on the RAI and always match
its predictions (i.e., a case of automation bias), most of the mass of the
boxplots would lie around 100\%. We observe that this does not occur. The median
influence of the RAI is always below 50\% and decreases as the gap between the
RAI's and the pre-registered risk estimates
decreases.}\label{fig:matrix_revision}
\end{figure}

As we discussed in \textsection\ref{sec:mapping_prob_bin}, participants' revised
risk estimates were closer to the RAI's than their pre-registered estimates.
Interestingly, the RAI's influence was 36\% higher when the participant's
pre-registered risk estimate was lower than the RAI's (mean inf. when
$Q^p_P<Q^p_{RAI}:\;44.8\%$, when $Q^p_{P}>Q^p_{RAI}:\; 32.9\%$; $p_{MW}<\alpha$)
but the average size of the revision was lower (mean $|Q^p_{P+RAI}-Q^p_{P}|$
when $Q^p_P<Q^p_{RAI}:\;6.7\%$, when $Q^p_P>Q^p_{RAI}: \;7.9\%$). The magnitude
of these revisions was not always orthogonal to the magnitude of the initial
difference: Influence was uncorrelated with the difference only in case of the
pre-registered probability prediction being lower than the RAI's ($\rho_S=0.2$
for $Q^p_P>Q^p_{RAI}$ and $p<\alpha$; $\rho_S=-0.05$ for $Q^p_P<Q^p_{RAI}$ and
$p > \alpha$) (see also Figure~\ref{fig:matrix_revision}
in~\textsection\ref{sec:additional_data}). The two results together indicate
that the relative---but not the absolute--- magnitude of participants' revisions
was larger in cases where they had initially underestimated the probability of
re-arrest compared to the RAI. This could be expected given that the initial
risk estimates made by participants generally were substantially higher than
those of the RAI (mean $|Q^p_P-Q^p_{RAI}|$ when $Q^p_P>Q^p_{RAI}$: $21.6\%$ vs
when $Q^p_P<Q^p_{RAI}$: $14.1\%$). Despite quite large heterogeneity in the
influence of the RAI across participants
(%($p_{KW}<\alpha$ for the comparison of mean influence across participants,
see Figure~\ref{fig:analysis_change_influence}),
we identified three notable patterns in their answers. 
Approximately 6\% of the participants always matched the RAI's predictions (17 out of 271), 
7\% never revised their risk estimates (19), 
and 11\% always averaged their initial risk estimates with the RAI's (29).
%We also found that the influence did not vary with the order of the question ($p_{KW}>\alpha$). 

Participants revised their pre-registered binary predictions in 10.2\% (se=0.3\%) of the assessments, 
switching in slightly more than half of these cases 
from a prediction of re-arrest to one of no re-arrest
(share of predictions that switch from re-arrest to no re-arrest: $5.7\%$).
% cite 
Here, two results are worth mentioning.
% first: participants often switch to the "wrong" prediction
First, when the participants changed their predictions, 
they not always did so to match the RAI's: 
This occurred only in 68.3\% of all revisions.
In 5.2\% of the cases where the participants' 
pre-registered answers matched the RAI's, 
their revised predictions did not.
% second: they are more likely to switch to a re-arrest
Second, the likelihood that participants would switch their prediction to match the RAI's varied with the actual recommendation: 
In cases where the pre-registered prediction did not correspond to the RAI's,
participants matched its prediction 49.8\% of the times when it was of a re-arrest, 
but only 14.8\% of the times when it was of no re-arrest ($p_{TT}<\alpha$, see also Figure~\ref{fig:no_anchoring} and Figure~\ref{fig:matrix_revision} in~\textsection\ref{sec:additional_data}).

These results might not seem surprising given our findings around how
participants converted probabilities into binary predictions in
\textsection\ref{sec:mapping_prob_bin}. Yet, unexpectedly, we found that the
direction of the difference between the pre-registered and the RAI's risk
estimates could not fully explain the direction of the revision. In cases where
the final answer matched the RAI's but the pre-registered prediction did not,
participants switched from a prediction of no re-arrest to one of re-arrest even
when the risk estimate of the RAI was {\it lower} than their pre-registered risk
estimate (47.5\% of 303 cases). Instead, when they switched from no re-arrest to
re-arrest---which also corresponded to the RAI's prediction---their
pre-registered risk estimate was almost always higher than the RAI's (89.2\% of
388 cases). 
%In cases where there was initial disagreement but then the participant matched
%the RAI's answer, the final prediction of no re-arrest occurred almost
%exclusively when the participants' risk estimate was lower than the RAI's
%(89.2\% of the cases); but in the same cases of initial disagreement
%participants' final predictions often corresponded to re-arrest even when their
%pre-registered risk estimates were lower than the RAI's! % such a complicated
%argument
In the right panel of Figure~\ref{fig:no_anchoring}, we observe that even when
the risk of re-arrest estimated by the RAI was very low, participants appeared
to be only slightly more likely to switch to a prediction of no re-arrest
($\rho_S=-0.07$ between $Q^p_{RAI}$ and $Q^b_{P}$ for $Q_{RAI}<50\%$;
$p<\alpha$). Lastly, we found that the risk
estimates of participants that self reported trust in the RAI (79.3\%) were
closer to the RAI's and that the RAI exerted higher influence on this set of
participants (mean influence if trust=$42.2\%$ otherwise=$24.7\%$;
$p_{MW}<\alpha$). The self-reported use of the RAI's risk estimates was weakly
correlated with the RAI's influence ($\rho_S=0.22$, $p<\alpha$).

\section{Structure and content of survey}\label{sec:app_surveystructure} 
\begin{figure}[!b]
  \centering
  \includegraphics[width=.33\linewidth]{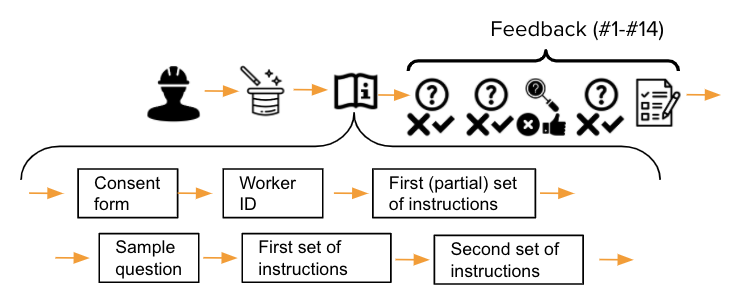}
\caption{Structure of the initial part of the survey. }\label{fig:instructions_zoomingin}
\end{figure}

In this
section we present the structure and content of the survey. We have adapted some
of the terminology used in the original survey to the one adopted in this
article. For example, the first and second part of the survey were called
``training'' and ``testing'' in the original survey. $Q^p$ and $Q^b$ were called
$Q1$ and $Q2$. Figure~\ref{fig:instructions_zoomingin} shows the structure of the
initial pages of the survey.

\subsection{Consent}\label{sec:app_consent}
\begin{center}
\fbox{
\parbox[H]{0.9\textwidth}{
\textbf{Purpose of research Study:}
The purpose of this study is to analyze how Amazon Mechanical Turk users interpret and use algorithmic predictions generated by decision support systems. We will perform statistical analysis to identify and extract patterns in the answers to this survey.\\

\textbf{Procedures:}
Your task is described in the following page.\\

\textbf{Risks/Discomforts:} 
There are no risks for participating in this study beyond those associated with
normal computer use and a minor risk of breach of confidentiality.\\

\textbf{Benefits:}
By completing this task, you will be more familiar with decision support systems used in the criminal justice setting. More broadly, this study may benefit society by improving the understanding of human-computer interaction.\\

\textbf{Voluntary participation and right to withdraw:}
Participation in this study is voluntary, and you can stop at any time. However, the survey needs to be completed for the HIT to be rewarded.\\

\textbf{Circumstances that could lead us to end your participation:}
We may decide to end your participation and/or not to reward the HIT for one of the following circumstances: (1) you fail at least one of the three attention checks (details in the instructions); (2) there is clear evidence that the questions have not been appropriately read and/or answered; (3) not all questions have been answered; (4) the survey has already been completed once, i.e. you are completing the survey for the second (or more) time.\\

\textbf{Confidentiality:}
Other than your Amazon Mechanical Turk serial number, we will collect the following demographic information: age, gender, race, level of education, and state of residence. We note that the Amazon Mechanical Turk serial number could be linked to your public profile page, so you might consider what information you choose to share on your public profile. These serial numbers will not be shared with anyone outside the research team and will only be used to handle financial transactions on the platform. Note, however, that de-identified data may be shared outside the research team.\\

\textbf{Compensation:}
If you satisfactorily complete the HIT/survey, you will receive a minimum compensation of 1.5\$ for your participation. The extra reward (up to 5\$) depends on your performance as described in the instructions. Payments are made via Amazon’s payment system.\\

\textbf{Contact information:}
[anonymized for submission]\\
%If you have any questions about this research, you may contact Riccardo Fogliato at rfogliat@andrew.cmu.edu. If you have any questions about your rights as a participant in this study or any concerns or complaints, please contact Carnegie Mellon University Institutional Review Board (IRB) Office at irb‐review@andrew.cmu.edu. We will also report any adverse events to the IRB Office in a timely manner, i.e. within 3 business days.
\textbf{Clicking accept:}
By clicking the “Accept” button, you indicate that you are 18 years of age or older, that you voluntarily agree to participate in this study, and that you understand the information in this consent form. You have not waived any legal rights you otherwise would have as a participant in a research study.\\
\centerline{\colorbox{gray!20}{\framebox(40,10){Accept}}}
}}
\end{center}

\subsection{Obtaining Worker ID}\label{sec:app_workerid}
\begin{center}
\fbox{
\parbox[H]{0.5\textwidth}{
Please enter your Amazon Mechanical Turk Worker ID. We will use it only to process the payment. After entering the code, please press enter. Remember that if you have already completed this task in the past, you will not be rewarded.
\centerline{\framebox(200,10){\texttt{Insert your ID}}}
\centerline{\colorbox{gray!20}{\framebox(70,10){Check if ID is new}}}
\centerline{\colorbox{gray!20}{\framebox(40,10){Continue}}}
}}
\end{center}

If the ID already existed in our database (i.e., the worker had either completed
the survey or failed an attention check in the past), the ``Continue'' button
would not appear. If the ID was modified after having been verified by clicking
the ``Check if ID is new'' button, then the ``Continue'' button was disabled.

\subsection{First (partial) set of instructions}\label{sec:app_partial_instructions}
\begin{center}
\fbox{
\parbox[H]{0.9\textwidth}{
\textbf{Task:} In this task you will be shown demographic information and criminal history for individuals that have been arrested. These offenders have all been selected from real data. For every offender, we know whether they were rearrested within three years of release. Based on the available information, your task will be to predict whether the offenders were rearrested and to estimate the likelihood of this event.\\

\textbf{Questions: }For each offender, you will be asked two questions:
\begin{itemize}
    \item $Q^p$: What is the likelihood of this offender being rearrested in the three years following release? For instance, 13\% means that you think the probability of rearrest is 13\%. You will use a slider to select your choice. The initial value of the slider is initialized at either 0 or 100.
    \item $Q^b$: Do you think that the offender was rearrested in the three years following release? This question is similar to the first but asks only for a yes-no prediction. 
\end{itemize}
It is important to answer $Q^p$ before $Q^b$. Once you have answered $Q^b$, you may not be able to revise your answer in $Q^p$.\\

In the three years following release, 42\% of all offenders in the population were rearrested. The set of offenders that is shown to you is a representative sample of the entire population. You will evaluate 40 different offenders (+ 1 example and +3 attention checks). Throughout the survey, we will ask you a short series of questions to make you reflect on your answers.\\

\begin{flushleft}Now you will be shown an example. Please answer the questions in the example before proceeding to read further instructions.\end{flushleft}
\centerline{\colorbox{gray!20}{\framebox(40,10){Continue}}}
}}
\end{center}

\subsection{First set of instructions}\label{sec:app_instructions} The first set
of instructions included what had already been presented
in~\textsection\ref{sec:app_partial_instructions}, plus the following panel.
Note that the third part of the survey (one question) was not mentioned in the
instructions.

\begin{center}
\begin{adjustbox}{width=0.9\textwidth}
\fbox{
\parbox[H]{\textwidth}{
\textbf{Structure of the survey:} The survey consists of two consecutive parts:
\begin{itemize}
    \item {\it First part (feedback, no algorithmic tool):} You will start the survey with some example cases. For each case, after making your prediction, you will receive feedback on whether the offender was rearrested within three years of release. The feedback indicates whether your prediction (in response to $Q^b$) was accurate. If you predicted that the offender was not rearrested, but in fact they were, then your prediction was inaccurate. If you predicted that the offender was rearrested, but in fact they were not, then your prediction was inaccurate. This feedback can also help you refine your answers to $Q^p$. If you find that you are often assigning high probabilities of re-arrest to offenders that are not rearrested, then these probabilities may be too high.
    \item {\it Second part (no feedback, algorithmic tool):} In this phase, you will make predictions but will not receive feedback about whether your predictions are accurate. For the questions in this part of the survey, you may also be shown the predictions of an algorithmic tool that was trained to predict re-arrest. You can incorporate the algorithmic tool’s predictions as an aid to help you evaluate, adjust, and revise your decisions. The tool is described below.\\
\end{itemize}

\textbf{Algorithmic tool:} 
The statistical model on which the tool is based was selected for being the most accurate among many statistical models. The tool has the following characteristics:
\begin{itemize}
    \item {\it Target:} The algorithmic tool has been trained to predict the likelihood of the offender’s re-arrest in the three years following release. For instance, if the algorithmic tool predicts that the likelihood of re-arrest is 60\% for some offender, it means that this offender is going to be rearrested with probability 60\%, according to this tool.
    \item {\it Information:} The algorithmic tool has access to all the information that is available to you. In addition, the tool has also access to more detailed categories of past crimes and their frequency.
    \item {\it Calibration:} Calibration is one of many benchmarks used to measure the quality of the predictions generated by algorithmic tools. Informally, an algorithmic tool is calibrated if on all the cases where it predicts, say, a 60\% probability of re-arrest, in fact 60\% of those offenders were rearrested. That does not mean that the algorithmic tool is actually good at making predictions. For example, an algorithmic tool that predicts that all offenders have a 42\% probability of re-arrest would be perfectly calibrated! Indeed, these are group level estimates rather than precise estimates of individual likelihoods of re-arrest. The calibration properties of this tool are shown in the figure below.\\
\end{itemize}
\begin{center}
\includegraphics[width = 0.5\textwidth]{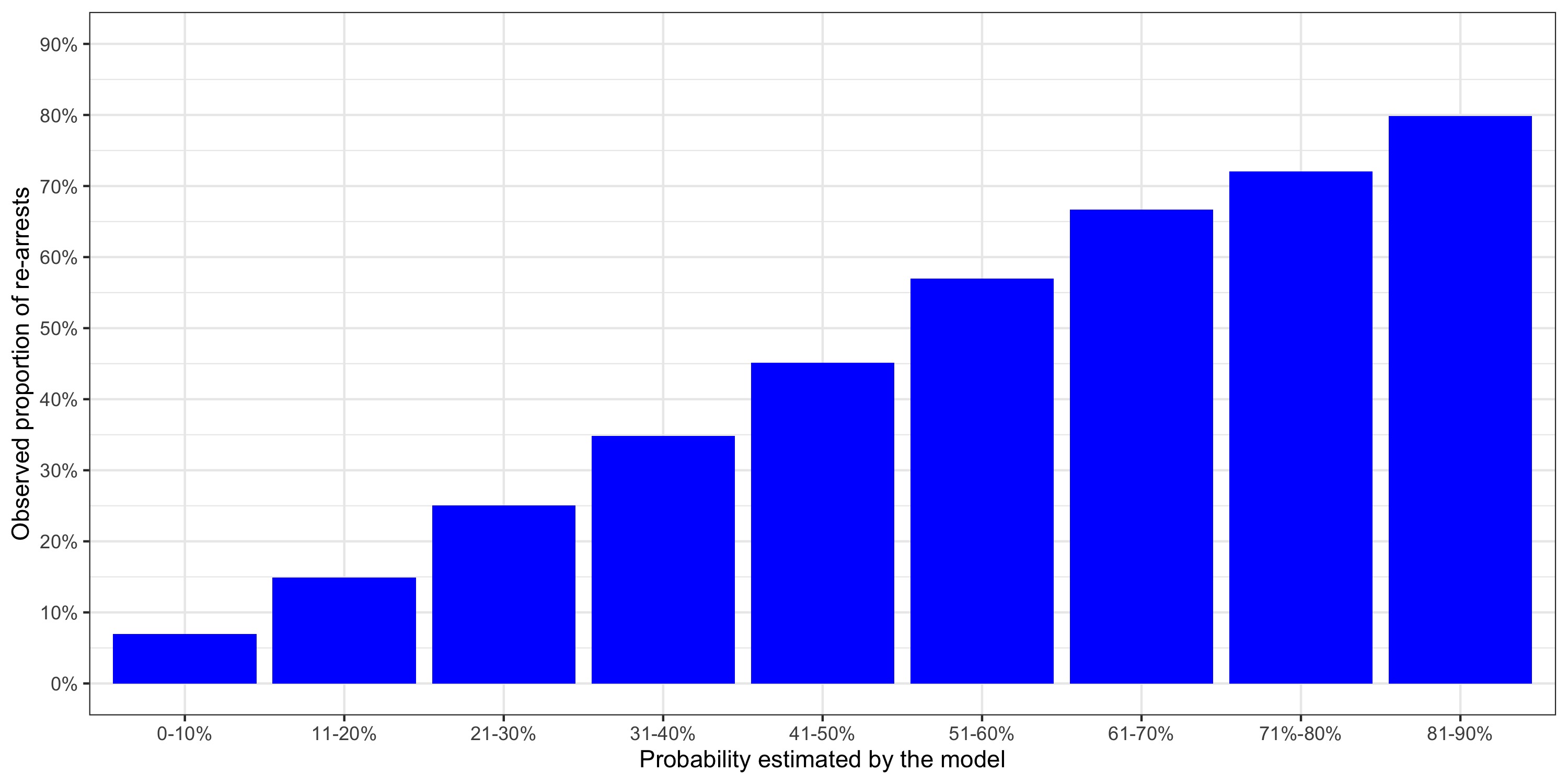}
\end{center}

\textbf{Terminology:} 
The following comparisons may be useful: \begin{itemize}
    \item {\it Misdemeanor vs felony:} A misdemeanor is a crime considered to be of lower seriousness compared to a felony. Misdemeanors carry up to 1 year of jail in most states. Notice that misdemeanors are more serious than infractions.
    \item {\it Adult vs juvenile} offenders under age 18 are considered juvenile.
\end{itemize}
\centerline{\colorbox{gray!20}{\framebox(40,10){Continue}}}
}}
\end{adjustbox}
\end{center}

\subsection{Second set of instructions}\label{sec:app_mechanism}
\begin{center}
\fbox{
\parbox[H]{0.95\textwidth}{
\textbf{Requirements:} You must answer all questions for the survey to be valid and to complete the HIT. For each page, please answer all questions before proceeding to the following page. You will also need to answer the attention checks (described below) correctly.\\

\textbf{Attention checks:} For three of the offenders, the answer to $Q^b$ will be explicitly mentioned in the description of the offender. You need to answer $Q^b$ accordingly. For those two offenders, you can select anything in $Q^p$. If you do not provide the correct answer to $Q^b$ (as specified in the text) for any of these two offenders, you will not be able to proceed with the rest of the survey. You will also not be able to retake the survey. If that's the case, please return the HIT.\\

\textbf{Structure of rewards:} The extra reward (up to 5\$) is based on your performance in the testing phase. Performance is proportional to the accuracy of your answers in $Q^p$ and $Q^b$. For this evaluation, we will use your answers to $Q^p$ for 13 randomly chosen offenders (out of 25), and your answers to $Q^b$ for the remaining offenders. The measurements of performance on $Q^p$ and $Q^b$ use Brier score and accuracy respectively. The maximum reward for each question is 0.20\$. For example, consider the case where you answer 30\% ($Q^p$) and "no" ($Q^b$), and the offender is not rearrested. In case $Q^p$ is selected, then you will receive 0.18\$ = 0.20\$*(1-(30\%)$^2$). In case $Q^b$ is selected, then you will receive the full 0.20\$. Conversely, if the offender is rearrested, then you will receive 0.10\$ in case of $Q^p$ and 0\$ in case of $Q^b$. The payment schema has been designed to ensure that you will achieve the highest reward only by acting according to your true beliefs. In other words, you will get the highest reward answering both answers as well as you can.

\centerline{\colorbox{gray!20}{\framebox(40,10){Continue}}}
}}
\end{center}

\subsection{Structure of the vignette}\label{sec:structure_vignette}

\begin{figure}[H]
\begin{center}
\fbox{
\parbox[H]{\textwidth}{
    {\small {\color{blue} \underline{Show/hide informed consent form and instructions}}}\\
    
    \underline{\textbf{Offender [\# OFFENDER] of 40}}\\
    
	{\bf Demographics:} The offender is [RACE] and [SEX]. She/he is [AGE] years old.\\
	
	{\bf Current charge:} The offender has been charged with [CRIME CATEGORY].\\
	
	{\bf Criminal history:} The offender (\texttt{if PRIOR ARRESTS>0})\{ has been arrested [PRIOR ARRESTS] time(s) and has been charged [PRIOR CHARGES] time(s). Of these arrests, (\texttt{if PRIOR ARRESTS-JUVENILE>0})\{[PRIOR ARRESTS-JUVENILE] occurred when (s)he was a juvenile.\} if(\texttt{if PRIOR ARRESTS-JUVENILE=0})\{all occurred when she/he was an adult.\} The previous charge was/charges were for the following categories of offenses: (\texttt{if PRIOR CHARGES-TRAFFIC>0})\{traffic\}, (\texttt{if PRIOR CHARGES-DRUGS>0})\{drugs\}, (\texttt{if PRIOR CHARGES-SEXUAL ASSAULT>0})\{sexual assault\}, (\texttt{if PRIOR CHARGES-PUBLIC ORDER>0})\{public order\}, (\texttt{if PRIOR CHARGES-PUBLIC ADMINISTRATION>0})\{public administration\}, (\texttt{if PRIOR CHARGES-PROPERTY>0})\{property\}, (\texttt{if PRIOR CHARGES-WEAPONS>0})\{weapons\}. Of these, [PRIOR CHARGES-VIOLENT] charge was/charges were for violent offenses.\} (\texttt{if PRIOR ARRESTS=0})\{has never been arrested before.\}\\
%\begin{itemize}
%	    \item[] $\ast$ \texttt{if PRIOR ARRESTS>0} $\cdot$ has been arrested [PRIOR ARRESTS] time(s) and has been charged [PRIOR CHARGES] time(s). Of these arrests,
%	    \begin{itemize}
%	        \item[] $\ast$ \texttt{if PRIOR ARRESTS-JUVENILE>0} $\cdot$ [PRIOR ARRESTS-JUVENILE] occurred when (s)he was a juvenile.
%	        \item[] $\ast$ \texttt{if PRIOR ARRESTS-JUVENILE=0} $\cdot$ all occurred when (s)he was an adult.
%	    \end{itemize}
%	    The previous charges were for the following categories of offenses: $\ast$ \texttt{if PRIOR CHARGES-TRAFFIC>0}$\cdot$ traffic, $\ast$\texttt{if PRIOR CHARGES-DRUGS>0}$\cdot$ drugs, $\ast$\texttt{if PRIOR CHARGES-SEXUAL ASSAULT>0}$\cdot$ sexual assault, $\ast$\texttt{if PRIOR CHARGES-PUBLIC ORDER>0}$\cdot$ public order, $\ast$\texttt{if PRIOR CHARGES-PUBLIC ADMINISTRATION>0}$\cdot$ public administration, $\ast$\texttt{if PRIOR CHARGES-PROPERTY>0}$\cdot$ property, $\ast$\texttt{if PRIOR CHARGES-WEAPONS>0}$\cdot$ weapons. Of these, [PRIOR CHARGES-VIOLENT] charge(s) was/were for violent offenses. 
%	    \item[] $\ast$\texttt{if PRIOR ARRESTS=0}$\cdot$ has never been arrested before.\\
%	\end{itemize}

	%$\ast$\texttt{if RAI shown}$\cdot$ The algorithmic tool estimates that the offender's likelihood of re-arrest is [$Q_{RAI}$].\\
	(\texttt{if RAI shown})\{The algorithmic tool estimates that the offender's likelihood of re-arrest is [$Q^p_{RAI}$].\}\\
	
	(\texttt{if RAI shown and non-anchoring setting})\{You have previously estimated that the offender's likelihood of re-arrest is [$Q_{P}^p$]. You have previously estimated that the offender WAS (\texttt{if $Q_{P}^b$=No})\{NOT\} %\begin{itemize}\item[] $\cdot$\texttt{if $Q_{P}^b$=No}$\cdot$ NOT \end{itemize} 
	going to be rearrested.\}\\
	
	\centerline{{\bf What is the likelihood of this offender being rearrested}} \centerline{{\bf in the three years following release?}}
	\begin{center}
	\includegraphics[width=0.7\textwidth]{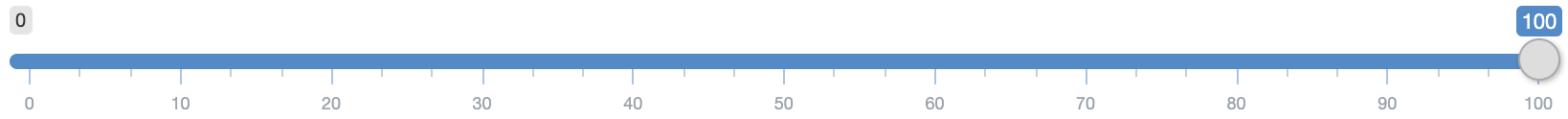}
	\end{center}
	\centerline{{\bf Do you think that the offender was rearrested}} \centerline{{\bf in the three years following release?}}
	\begin{center}
	\includegraphics[width=0.1\textwidth]{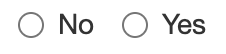}
	\end{center}
	\noindent (\texttt{if both answers given}) \qquad\qquad\qquad  \includegraphics[width=0.1\textwidth]{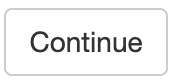}\\
	(\texttt{if feedback shown})\{(\texttt{if $Y=Q_P^b$})\{{\color{green} The offender WAS (\texttt{if $Y$=No})\{NOT\} rearrested in the three years following release.}\} (\texttt{if $Y\neq Q_P^b$})\{{\color{red} The offender WAS (\texttt{if $Y$=No})\{NOT\} rearrested in the three years following release.}\}\}
	%\begin{itemize}
	 %   \item[] \texttt{if $Y=Q_P^b$}$\cdot$ {\color{green} The offender WAS (\texttt{if $Y$=No})\{NOT rearrested in the three years following release.\}}
	    %\begin{itemize}
	    %    \item[] $\ast$\texttt{if $Y$=No}$\cdot$ {\color{green} NOT}
	    %\end{itemize}{\color{green} rearrested in the three years following release.}
	  %  \item[]$ \ast$\texttt{if $Y\neq Q_P^b$}$\cdot$ {\color{red} The offender WAS (NOT) rearrested in the three years following release.}
	%\end{itemize}
	
	}
}
\end{center}
\caption{Structure of the vignette. If the condition inside round brackets was met, then the value taken by the feature inside curly brackets was inserted into the text. For example, if \texttt{PRIOR ARRESTS} was ``5'', then the generated sentence was ``The offender has been arrested 5 times...''. $Q_P^p$ and $Q_P^b$ indicate the answers given by the participant without the assistance of the RAI to the first and second question in the vignette respectively. $Y$ indicates the outcome of the offender (i.e., rearrested or not). The RAIs presented in our work were trained only on the features, such as the offender's prior number of arrests and age, and not on the full sentences.}\label{fig:structure_vignette}
\end{figure}

\subsection{Demographics}\label{sec:app_demo}
The participant is allowed to proceed only once all questions have been answered.
\begin{center}
\begin{adjustbox}{width=0.7\textwidth}
\fbox{
\parbox[H]{0.9\textwidth}{
\begin{itemize}
    \item How old are you?
    \begin{itemize}
    \item {[Integers from 18 to 80 to be selected using a slide bar]}
    \end{itemize}
    \item What's your gender? \begin{itemize}[label=$\circ$]
        \item Female
        \item Male
        \item Transgender female
        \item Transgender male
        \item Not in these categories
        \item Prefer not to say
    \end{itemize}
    \item What's your race/ethnicity?
    \begin{itemize}[label=$\circ$]
        \item White
        \item Black or African American
        \item American Indian or Alaska Native
        \item Asian
        \item Native Hawaiian or Other Pacific Islander
        \item Not in these categories
        \item Prefer not to say
    \end{itemize}
    \item What's your highest level of education? (already achieved)
       \begin{itemize}[label=$\circ$]
        \item Less than high school degree
        \item Some college but no degree
        \item Associate degree
        \item Bachelor degree
        \item Graduate degree
        \item Not in these categories
        \item Prefer not to say
    \end{itemize}
    \item Where do you leave?
    \begin{itemize}
        \item {[Any of the US states to be selected using a drop-down list]}
    \end{itemize}
    \item Have you ever used (even on Amazon MTurk) decision support systems (algorithmic tools) like the one described in the instructions?
    \begin{itemize}[label=$\circ$]
        \item No
        \item Yes
    \end{itemize}
\end{itemize}
\centerline{\colorbox{gray!20}{\framebox(40,10){Continue}}}
}
}
\end{adjustbox}
\end{center}

\subsection{Perception questions}\label{sec:app_perception_questions} The
participant was allowed to proceed with the rest of the survey only once all questions other
than the two free-response questions had been answered. Two of the questions in the second set of perception questions were
affected by a technical issue in the initial batches of surveys. 
% There, we asked
% participants to evaluate their own performance and reliance on the tool for the
% 14 offenders presented in that part of the survey. But participants had seen only
% 13 offenders by that point. For this reason, we omitted part of these questions
% from the analysis.
For the purpose of the data analysis, in question (1) we coded ``Not confident
at all'' and ``Slightly confident'' as ``Not confident'', ``Somewhat
confident'' as ``Neutral'', and the remaining two categories as
``Confident''. In question (4), we coded the lowest two levels as ``Below median
accuracy'', the middle level as ``Around median accuracy'', and the remaining
two categories as ``Above median accuracy''.
\begin{center}
\begin{adjustbox}{width=0.8\textwidth}
\fbox{\parbox[H]{\textwidth}{You have already evaluated [percentage of offenders
already evaluated]\% of the offenders! Please answer the following questions
very carefully. \\
\noindent (\texttt{if second part of the survey})\{These questions concern all answers given in the testing phase of the survey. Do not take into account the decisions that you made during the first part of the survey.\}\\
(\texttt{if second part of the survey and non-anchoring setting})\{Consider only those predictions made after taking into account the algorithmic tool's prediction.\}
\begin{enumerate}
    \item How confident were you, on average, in your yes-no predictions? 
           \begin{itemize}[label=$\circ$]
        \item Not confident at all
        \item Slightly confident 
        \item Somewhat confident 
        \item Moderately confident
        \item Extremely confident
    \end{itemize}
    \item If you happened to be very confident in some of your yes-no predictions, which characteristics of the offender made you so confident? Note that you may be not very confident on average, but still be extremely confident in some of your predictions.\\
    \framebox(200,10){\texttt{Insert your comments here}}
    \item How many of your yes-no predictions do you think were correct?
    \begin{itemize}
        \item {[integers from 0 to 13, 14 or 25 to be selected using radio buttons]}
    \end{itemize}
    \item How do you think that the accuracy of your yes-no predictions
            compares to the accuracies of other Amazon Mechanical Turk workers in this task? \\The percentages in parentheses refer to the percentile rank of your accuracy
            in the distribution of all workers\' accuracies. For example, say that there are other 100 workers.
            If you think that your accuracy is higher than the accuracies of 35 of them, then you should choose the percentile range (21-40\%).
            Instead, if you think that your accuracy is higher than the accuracies of 85 of them, then you should choose the percentile range (81-100\%).
    \begin{itemize}[label=$\circ$]
        \item Among the lowest accuracies (0-20\%)
        \item Lower than most accuracies (21-40\%)
        \item Approximately equal to the median accuracy (41-60\%)
        \item Higher than most accuracies (61-80\%)
        \item Among the highest accuracies (81-100\%)
    \end{itemize}
    \item (\texttt{if second part of the survey})\{How often did you revise (change) your {\it numerical} prediction after seeing the prediction of the algorithmic tool?\}
    \begin{itemize}
        \item {[integers from 0 to 13, 14, or 25 to be selected using radio buttons]}
    \end{itemize}
    \item (\texttt{if second part of the survey})\{How often did you revise (change) your {\it yes-no} prediction after seeing the prediction of the algorithmic tool?\}
    \begin{itemize}
        \item {[integers from 0 to 13, 14, or 25 to be selected using radio buttons]}
    \end{itemize}
    \item (\texttt{if second part of the survey})\{If some of your yes-no predictions did not match the algorithmic tool\'s and you decided not to  revise your yes-no predictions, why did you think that you were more accurate than the algorithmic tool?\}\\
    \framebox(200,10){\texttt{Insert your comments here}}
\end{enumerate}
\centerline{\colorbox{gray!20}{\framebox(40,10){Continue}}}
}}\end{adjustbox}\end{center}

\section{Additional tables and figures}\label{sec:additional_data}

\begin{table}[H]
\begin{tabular}{ccll}
\toprule
       {\bf Category}                      &      {\bf Value}                           & {\bf Census} & {\bf Survey (n)}     \\
                            %\hline
                            \toprule
\multirow{2}{*}{Sex (Census) /Gender (survey)} & Male                                     & 49\%   & 62.5\% (332)     \\
                            & Female                                   & 51\%   & 36.5\% (194)      \\
                            \midrule
\multirow{4}{*}{Race}       & White                                    & 76\%   & 76.1\% (404)     \\
                            & Black or AA                              & 13\%   & 14.7\% (78)      \\
                            & Asian                                    & 6\%    & 6.2\% (33)       \\
                            & AI, AN, NH, PI                           & 2\%    & 1.7\% (9)        \\
                            \midrule
\multirow{4}{*}{Age}        & 18-24 years old                          & 7\%    & 2.8\% (15)       \\
                            & 25-34 years old                          & 10\%   & 44.1\% (234)      \\
                            & 35-59 years old                          & 25\%   & 46\% (244)      \\
                            & 60-78 years old                          & 19\%   & 6.8\% (36)       \\
                            \midrule
\multirow{1}{*}{Education}  & College degree or higher (age 25+)       & 32\%   & 81.2\% (419/516) \\
                            \midrule
\multirow{4}{*}{Region}     & Northeast                                & 17\%   & 16.9\% (90)      \\
                            & Midwest                                  & 21\%   &  19.1\% (102)   \\
                            & West                                     & 24\%   & 24.8\% (132)        \\
                            & South                                    & 38\%   & 39.0\% (207)      \\
                            \midrule
ML familiarity              & Yes                                      &        & 44\% (93) \\ \bottomrule
\end{tabular}
\caption{Comparison of the demographics of the sample in the survey with the estimates relative to the US population, obtained from the Census. The categories with only a few observations are excluded from the table. While we measure gender, the Census only provides estimates for sex.}\label{tab:demo}
\end{table}

\begin{table}[H]
\begin{tabular}{cccc}
\toprule
\multicolumn{4}{c}{{\bf Judges}}\\
  \toprule
{\bf Term} & {\bf Estimate} & {\bf Std.error} &  {\bf p.value} \\ 
 \midrule
 (Intercept) & -2.65 & 0.03 & 0.00 \\ 
 PRS & 0.38 & 0.01 & 0.00 \\ 
 OGS & 0.49 & 0.01 & 0.00 \\ 
   \bottomrule
\end{tabular}
\begin{tabular}{cccc}
\toprule
\multicolumn{4}{c}{{\bf RAI}}\\
  \toprule
{\bf Term} & {\bf Estimate} & {\bf Std.error} &  {\bf p.value} \\ 
 \midrule
 (Intercept) & -1.07 & 0.03 & 0.00 \\ 
 PRS & 0.42 & 0.01 & 0.00 \\ 
 OGS & 0.08 & 0.01 & 0.00 \\ 
   \bottomrule
\end{tabular}
\begin{tabular}{ccccc}
\toprule
\multicolumn{4}{c}{{\bf Participants}}\\
  \toprule
{\bf Term} & {\bf Estimate} & {\bf Std.error} &  {\bf p.value} \\ 
 \midrule
 (Intercept) & -0.23 & 0.03 & 0.00 \\ 
 PRS & 0.36 & 0.01 & 0.00 \\ 
 OGS & 0.05 & 0.01 & 0.00 \\ 
   \bottomrule
\end{tabular}
\quad
\caption{Summaries of logistic regression models targeting the likelihood of
incarceration (by judges) or the likelihood of predicted re-arrest (by RAI's and
participants' predictions). The tables contain coefficients estimates, sandwich
standard errors, and p-values relative to the null hypothesis that the
coefficients are equal to zero. These models have been described
in~\textsection\ref{sec:pred_vs_dec}. The coefficients correspond to
the offense gravity score (OGS) and to the prior record score (PRS). We note
that the coefficient's estimate of the offense gravity score is large in the model
targeting the likelihood of incarceration, but small in the others.}\label{tab:log_models}
\end{table}

\section{Comparison between the pilot study and the main experiment}\label{sec:comparison_pilot}

Before running the experiment described in the main paper, we conducted a pilot study. 
The survey used in this study was in many ways similar to the survey in the main experiment, but also differed in the following aspects: 
\begin{itemize}
    \item participants were all shown the same set of offenders but in different orders. The offenders' descriptions were based on a random sample drawn from the holdout set. The offenders were divided into two groups. The order was then randomized within groups.
    \item absence of bonus proportional to predictive performance and attention checks. In the pilot study, participants' reward was not tied to their performance. We paid a fixed amount of \$10 to all participants that completed the survey. We did not use attention checks.
    \item probability predictions were elicited using a slider instrument with a 10-point probability scale (0-10\%, 11-20\%, 21-30\%, etc.).
\end{itemize}
%In the pilot study, the decreasing and controlled conditions were not present. 
In the pilot study, 24 of the 50 participants were assigned to the anchoring
setting. Most of the findings from the pilot survey are virtually identical to
those that were presented for the main study. In particular, the results
of~\textsection\ref{sec:mapping_prob_bin} (probability and binary predictions)
and~\textsection\ref{sec:study_anc} (anchoring effects), which in principle
could have been affected by the choice of the probability scale, held in this
survey as well. We briefly discuss three results from the pilot study.

\begin{figure}[H]
\centering
\begin{subfigure}{.5\textwidth}
  \centering
  \includegraphics[width=\linewidth]{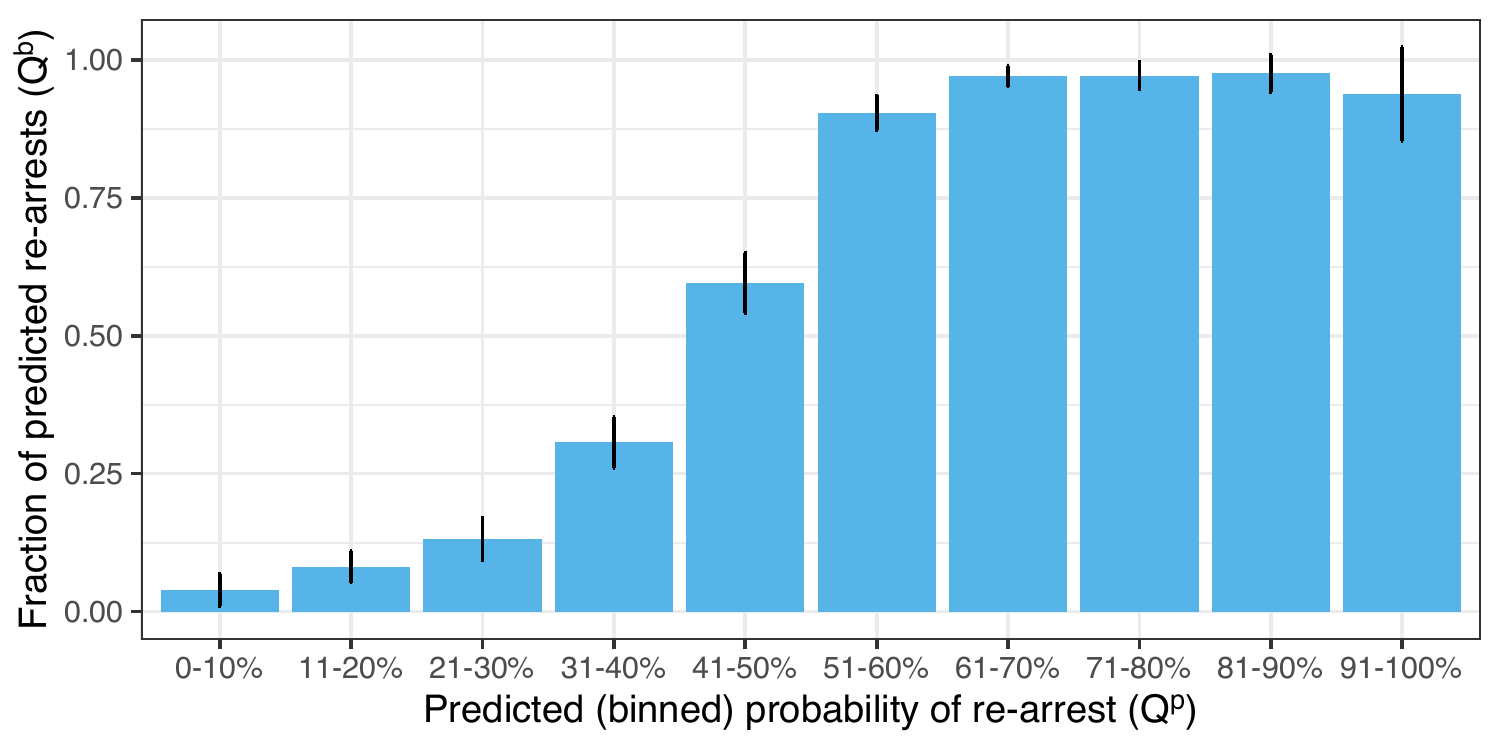}
\end{subfigure}%
\begin{subfigure}{.5\textwidth}
  \centering
  \includegraphics[width=\linewidth]{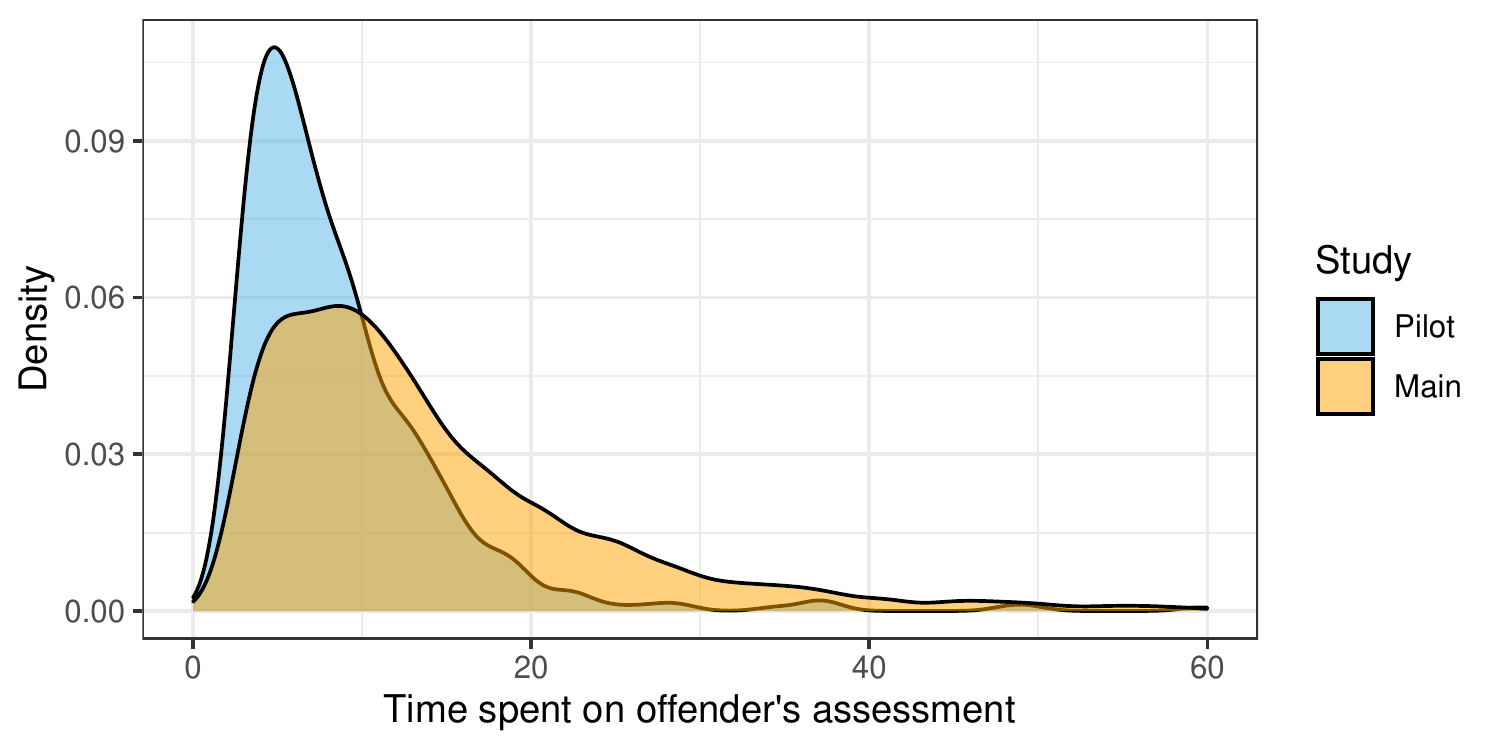}
 % \caption{}
\end{subfigure}%
\caption{Analysis of the participants' assessments in the pilot survey. (left)
Share of predictions of re-arrest as a function of the probability
predictions. Error bars indicate 95\% confidence intervals. As in the main
experiment (see Figure~\ref{fig:mapping_probs}), we observe that binary
predictions corresponding to re-arrest were frequent even for low values of the
likelihood. (right) Comparison of time spent on each offender's assessment in
the anchoring setting in the pilot and main study. Here, we considered only the
predictions made in presence of the RAI. These results indicate that assessments  in the main study
generally took longer than those in the pilot ($p_{MW}<\alpha$).}\label{fig:pilot_mapping}
\end{figure}

As in main study, we found that participants often predicted re-arrest even when
they were assigning low values to its likelihood. Figure~\ref{fig:pilot_mapping}
shows the share of predicted re-arrests as a function of the participants' risk
estimates. The pattern in these predictions closely resembles what has been
shown for the main experiment (see Figure~\ref{fig:mapping_probs}).
``Re-arrest'' was predicted for approximately one fourth of the offenders for
whom the risk estimate was between 41-50\% or lower (mean re-arrest
$Q^b$=26.4\%) but ``no re-arrest'' was predicted for only 5\% of the offenders
whose risk estimates were equal or above 51\%.

\begin{figure}[H]
\centering
\includegraphics[width=\linewidth]{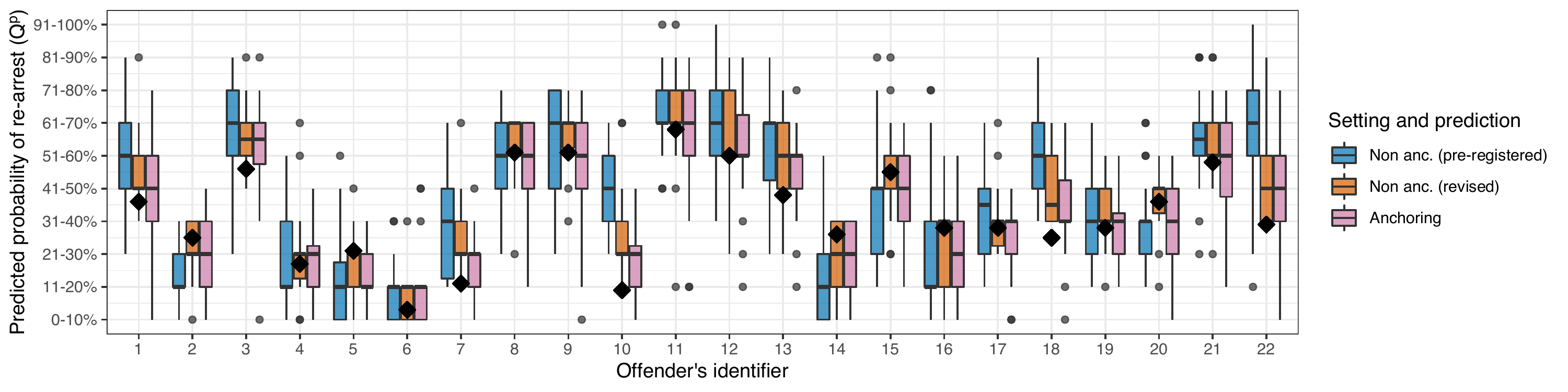}
\caption{Analysis of participants' and RAI's predictions in the pilot study.
Here, the anchoring effect was tested on a total of 22 different offenders. Each
of the offenders corresponds to a separate identifier on the horizontal axis.
The boxplots are relative to the pre-registered and revised risk estimates made in
non-anchoring setting, and those made in the anchoring setting. Each boxplot
corresponds to predictions made for a certain offender. The black squares
indicate the prediction of the RAI that was shown to the
participants.}\label{fig:anchoring_vs_not}
\end{figure}

As in the main study, we found that the revised risk estimates of the
participants in the non-anchoring setting were closer to the RAI's than those of
the participants assigned to the anchoring setting (mean
$|Q^p_{P+RAI}-Q^p_{RAI}|$ in non-anchoring=$0.09$, in anchoring=$0.12$;
$p_{TT}<\alpha$).\footnote{For each probability bin, we took its average value
(e.g., 11-20\% was converted into 15.5\%).} The agreement between participants'
and RAI's binary predictions, however, was slightly higher in
the anchoring setting (mean $|Q^b_{P+RAI}-Q^b_{RAI}|$ in non-anchoring=32\%, in
anchoring=27\%). Figure~\ref{fig:anchoring_vs_not} shows the distribution of the
probability predictions in the two settings for the 22 different offenders on
which we tested anchoring effects. We observe that participants assigned to the
non-anchoring setting often revised their risk estimates in the direction of the
RAI's. For most of the offenders the difference between the pre-registered
predictions and the RAI's estimates was fairly small. In cases where the
pre-registered risk estimates were far from the RAI's, the distribution of the
revised predictions was similar to the one of the predictions made in the
anchoring setting sometime (e.g., offender with id 22) but not in all cases (e.g., offenders with
id 7 and 10).
% speculation: you often change your probability prediction but not your binary answer
% speculation #2: people are less likely to use extreme scores with this scale

Since the vignette did not change, the time spent by participants on the
individual assessments can be compared across the two surveys. We separated the
predictions made without the assistance of the RAI from those made in presence
of the RAI. In case of the latter, we excluded those made in the non-anchoring
setting. For both groups of predictions, we found that participants in the main
study spent substantially more time than those in the pilot survey (both
$p_{MW}<\alpha$). The right panel of Figure~\ref{fig:pilot_mapping} shows the
distribution of time spent on each assessment in the anchoring setting: The
assessment took on average 16.1 seconds (median=11.4) for the participants in the
main study but only 9.7 seconds (median=7) for those of the pilot study. It is
possible that the introduction of attention checks, together with the incentive
structure being tied to predictions accuracy, nudged participants into spending
more time on the assessments or led to the exclusion of participants that were
not paying adequate attention.

\received{January 2021}
\received[revised]{April 2021}
\received[revised]{July 2021}
%% OR depending on your acceptance date, see linked additional info above
\received[accepted]{July 2021}

\end{document}